\documentclass[sigconf, authorversion]{acmart}

\usepackage{hyperref}
\graphicspath{{figs/}{figures/}{pictures/}{images/}{./}} 

\usepackage{tabu}                      
\usepackage{booktabs}                  
\usepackage{lipsum}                    
\usepackage{mwe}                       
\usepackage{amsmath}                   
\usepackage{newtxmath}  
\usepackage{svg}
\usepackage{graphicx}    
\usepackage{subcaption}
\usepackage{placeins}
\usepackage{caption}
\usepackage{xcolor}

\usepackage[normalem]{ulem}
\usepackage{amsmath}

\usepackage{submission}
\usepackage{macros}

\setmode{submission}
\AtBeginDocument{%
  \providecommand\BibTeX{{%
    \normalfont B\kern-0.5em{\scshape i\kern-0.25em b}\kern-0.8em\TeX}}}

\copyrightyear{2025}
\acmYear{2025}
\setcopyright{cc}
\setcctype{by}
\acmConference[VRST '25]{31st ACM Symposium on Virtual Reality Software and Technology}{November 12--14, 2025}{Montreal, QC, Canada}
\acmBooktitle{31st ACM Symposium on Virtual Reality Software and Technology (VRST '25), November 12--14, 2025, Montreal, QC, Canada}
\acmDOI{10.1145/3756884.3766004}
\acmISBN{979-8-4007-2118-2/2025/11}

\begin{document}


\title{Effects of Virtual Controller Representation and Virtuality on Selection Performance in Extended Reality}

\renewcommand{\shorttitle}{Effects of Visual Representations in VR and MR}

\author{Eric DeMarbre}
\affiliation{%
  \institution{Carleton University}
  \city{Ottawa}
  \country{Canada}}
\email{eric.demarbre@carleton.ca}

\author{Jay Henderson}
\affiliation{%
  \institution{Memorial University}
  \city{St. John's}
  \country{Canada}}
\email{jayhend@mun.ca}

\author{J. Felipe Gonzalez}
\affiliation{%
  \institution{Carleton University}
  \city{Ottawa}
  \country{Canada}}
\email{johannavila@cmail.carleton.ca}

\author{Robert J. Teather}
\affiliation{%
  \institution{Monash University}
  \city{Melbourne}
  \country{Australia}}
\email{rob.teather@monash.edu}

\renewcommand{\shortauthors}{E DeMarbre et al.}


\begin{abstract}


We present an experiment exploring how the controller's virtual representation impacts target acquisition performance across MR and VR contexts. Participants performed selection tasks comparing four visual configurations: a virtual controller, a virtual hand, both the controller and the hand, and neither representation. We found performance comparable between VR and MR, and switching between them did not impact the user's ability to perform basic tasks. Controller representations mimicking reality enhanced performance across both modes. However, users perceived performance differently in MR, indicating the need for unique MR design considerations, particularly regarding spatial awareness.

\end{abstract}



\begin{teaserfigure}
  \centering
  \includegraphics[width=0.23\textwidth]{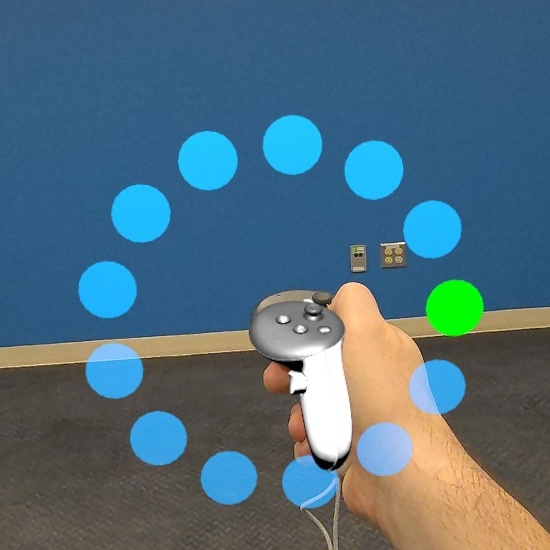}
  \includegraphics[width=0.23\textwidth]{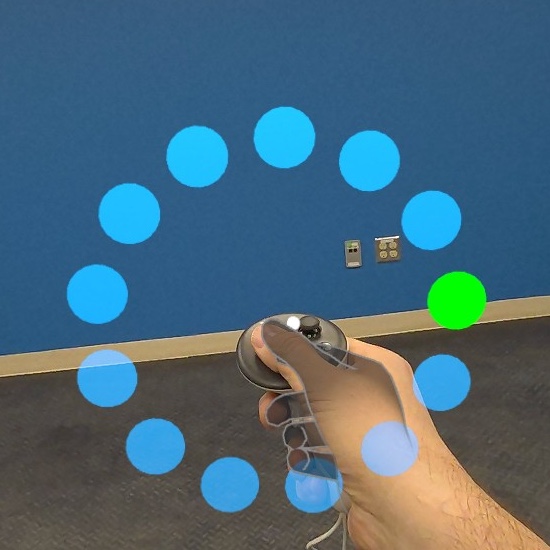}
  \includegraphics[width=0.23\textwidth]{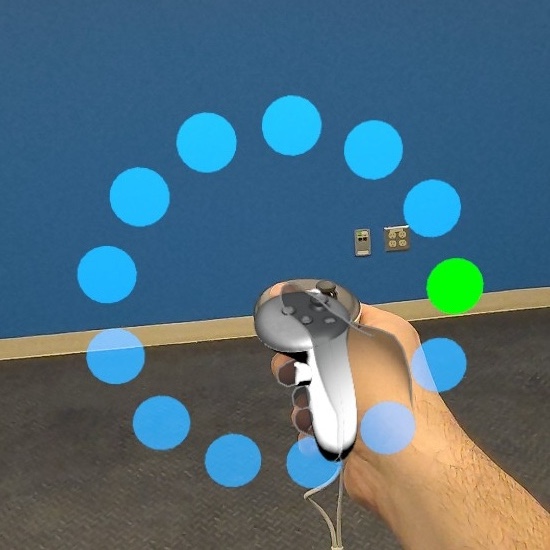}
  \includegraphics[width=0.23\textwidth]{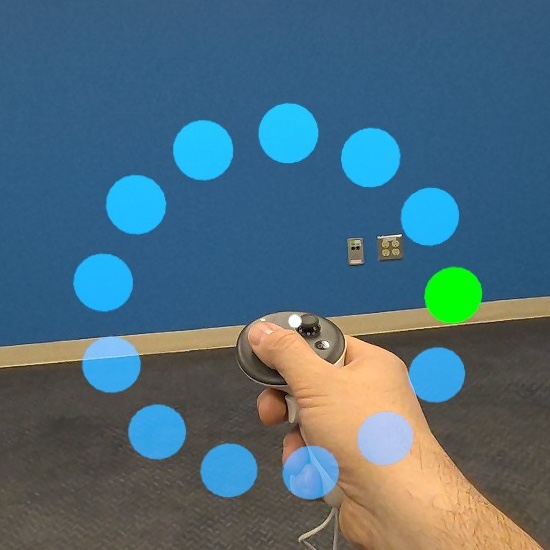}
  \vspace{1mm}
  \makebox[0.23\textwidth][c]{\scriptsize (a) MR – Controller}
  \makebox[0.23\textwidth][c]{\scriptsize (b) MR – Hand}
  \makebox[0.23\textwidth][c]{\scriptsize (c) MR – Both}
  \makebox[0.23\textwidth][c]{\scriptsize (d) MR – None}
  \vspace{1mm}
  \includegraphics[width=0.23\textwidth]{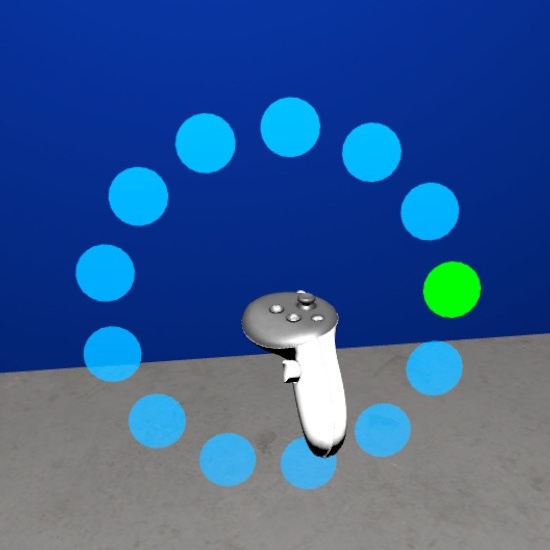}
  \includegraphics[width=0.23\textwidth]{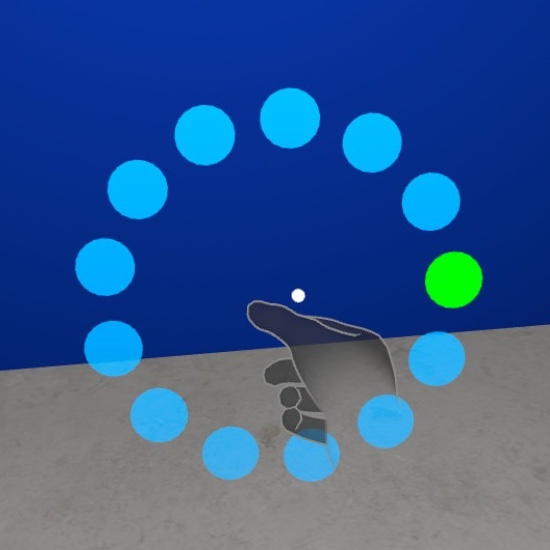}
  \includegraphics[width=0.23\textwidth]{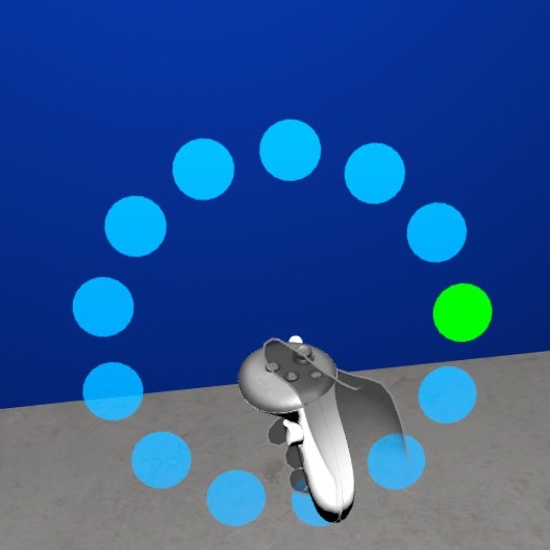}
  \includegraphics[width=0.23\textwidth]{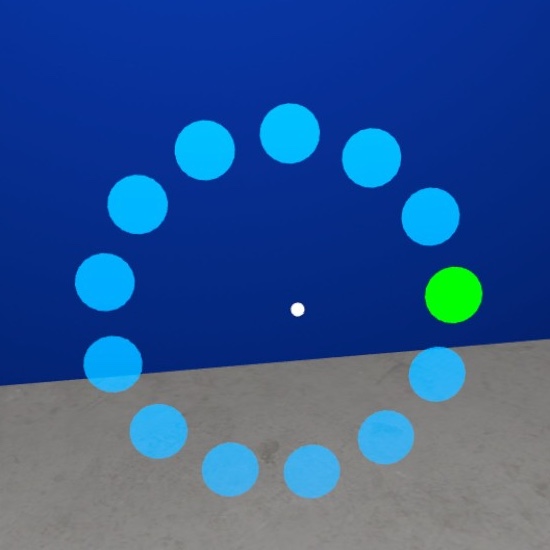}
  \vspace{1mm}
  \makebox[0.23\textwidth][c]{\scriptsize (e) VR – Controller}
  \makebox[0.23\textwidth][c]{\scriptsize (f) VR – Hand}
  \makebox[0.23\textwidth][c]{\scriptsize (g) VR – Both}
  \makebox[0.23\textwidth][c]{\scriptsize (h) VR – None}
  \caption{The 8 (4×2) experimental conditions with controller representation (left to right: controller only, hand only, both controller and hand, and no representation), across reality modes: Mixed Reality (top row) and Virtual Reality (bottom row).}
  \label{fig:conditions-no-task}
\end{teaserfigure}


\hyphenpenalty=9500  
\tolerance=1000

\maketitle

\section{Introduction}

Recent extended reality (XR) head-mounted displays (HMDs), such as the Meta Quest \cite{metaMetaQuestMR2025}, integrate cameras that capture the physical environment, enabling video see-through (VST) capabilities that merge real and virtual content~\cite{kress2017optical}. As a result, these devices can support applications across the entire \textit{Reality–Virtuality Continuum}~ \cite{milgram1994taxonomy, skarbez2021revisiting}. Moreover, users can dynamically switch between different stages of the spectrum, from augmented reality (AR) scenes, interactive scenes of virtual information in physical worlds with mixed reality (MR), to fully immersive virtual reality (VR) environments~\cite{pointeckerRealVirtualExploring2024, pointeckerBridgingGapRealities2022, liMixedRealityTunneling2022}.

Interactions are central to the user experience in VR and MR, as they enable effective engagement with virtual content and directly impact immersion, usability, and presence~\cite{myersInteractingDistanceMeasuring2002, kopper_human_2010}. Development frameworks typically offer tools like ray casting, direct touch, or hand gestures, employing hand-tracking or controllers \cite{metaMetaQuestMR2025}. These tools often display hands or controllers as visual representations to guide interaction. In VR, such representations are central to maintaining immersion and embodiment, especially as they serve as the user's only visual anchor \cite{seinfeldUserRepresentationsHumanComputer2021, lin_need_2016, hanashima_how_2022}. In MR, the use of avatar-based visual representations has been more limited, likely due to the continued visibility of the user’s real hands and controllers. However, some approaches have shown that augmenting the body with virtual elements, such as extended limbs or third-person views, can still enhance embodiment and interaction performance \cite{feuchtnerExtendingBodyInteraction2017, genayBeingAvatarReal2022, venkatakrishnanGiveMeHand2023}. 
To date, it remains unclear whether using similar avatar-based visual representations yields comparable user experiences and task performance in MR and VR, given the perceptual and contextual differences between the two environments.

One factor that may influence interaction differences between MR and VR is depth perception. In VR, issues like the conflict between vergence and accommodation and perspective scaling of targets at greater depths can impact pointing performance \cite{hoffman_vergenceaccommodation_2008, kovacs_perceptual_2008}. In MR, particularly with video see-through, depth perception issues can yield depth cue conflicts between virtual and real elements, leading to higher misjudgment rates and compensatory head movements \cite{westermeierAssessingDepthPerception2024}.

Interaction and visual representation function differently in MR compared to VR, as MR users must navigate the spatial alignment of their own perspective, real-world elements, and virtual content, relying heavily on physical cues like depth perception to make sense of these layers \cite{drascic1996perceptual, jones2008-effects, kruijff2010perceptual}. These differences in these mechanisms are not well understood, which poses challenges for applications aiming to support MR-VR experiences with consistent interaction design. For example, previous research has recreated real-world scenarios in VR and enabled transitions between MR-VR modalities \cite{westermeierAssessingDepthPerception2024, pointeckerRealVirtualExploring2024, demarbre_investigating_2024}. In such cases, replicating the same interaction techniques across platforms without accounting for the perceptual differences between VR and MR can negatively impact both user performance and experience.

In brief, there is still limited understanding of how different visual representations affect user performance and experience across VR and MR, and to what extent results obtained in one modality can be considered comparable to the other.


To address these gaps, we conducted a study to examine how visual representations affect selection performance in both VR and MR. Participants performed a Fitts’ law-based task under two XR modes, VR and MR, and four visual representation conditions: virtual controller, virtual hand, both combined, and no visual feedback.

Motivating our work are the following research questions:

\noindent
\textbf{RQ1}: Are there quantifiable differences in selection time, throughput, and depth error due to differences between virtual and mixed reality?

\noindent
\textbf{RQ2}: Are there quantifiable differences between selection time, throughput, and depth error due to variations in how the controller and hand are represented?

\noindent
\textbf{RQ3}: How do different controller representations influence user subjective preference?

Our principal contribution is a systematic evaluation of the effects of controller representation in target selection performance across VR and MR environments. Our findings inform the fundamentals of selection for developers, designers, and researchers of pass-through mixed reality applications.
\section{Previous work}

\subsection{Distinguishing Mixed and Virtual Reality}

The Reality–Virtuality Continuum refers to a spectrum of displayed environments that combine real and virtual content, with one end presenting a completely real environment and the other presenting a fully virtual environment, which is what we refer to as virtual reality (VR) \cite{milgram1994taxonomy}. Any combination not falling on the extremes is often referred to as \textit{Mixed Reality} (MR)
where physical environments are enhanced with interactive digital content. In this paper, we investigate how visual representation affects user experience and performance by comparing interactions in both VR and MR settings.

In this work, we focus on the two sides of the continuum where visual representation may have a significant impact on interaction design: VR, with fully embedded interactive virtual environments, and MR, where digital elements coexist interactively with physical space.

\subsection{Visual Representation}

The use of visual representations may be distinct between VR and MR. In VR, visual representation plays a crucial role in maintaining immersion and improving interaction fidelity \cite{ebrahimiInvestigatingEffectsAnthropomorphic2018, lougiakisEffectsVirtualHand2020, mcmanusInfluenceAvatarSelf2011}, especially since VR removes all external visual feedback and the avatar becomes the user's primary reference point \cite{seinfeldUserRepresentationsHumanComputer2021}. Visual representations—particularly of hands or controllers—help establish a meaningful link between users’ physical actions and their virtual counterparts. Research shows that accurate and congruent avatars enhance embodiment, realism, and performance—for example, matching hand shape and motion boosts embodiment \cite{lin_need_2016}. Similarly, aligning the visual avatar of a controller with its physical form enhances user control and reduces discomfort \cite{ponton_stretch_2024, hibbs_impact_2024}. Studies also highlight that visuo-motor-tactile synchrony enhances body ownership and realism during interaction \cite{hanashima_how_2022}. Moreover, matching the visual representation with the input technique can directly affect user performance and experience \cite{venkatakrishnan_how_2023, johnson_exploring_2023}. These effects are often linked to the user's sense of embodiment and presence \cite{kilteni_sense_2012, witmer_measuring_1998}, as even partial representations—such as virtual controllers—can influence proprioception and the feeling of ‘being there’ \cite{castronova2003theory, slater_body_1994}.

The uses of avatars in MR are different. Research has explored how avatar representations influence user experience and perception in mixed environments. For example, Genay \ea~\cite{genayBeingAvatarReal2022} examined third-person perspectives in AR, showing that changes in avatar appearance can affect self-perception, relating to the Proteus effect. Feuchtner and Müller \cite{feuchtnerExtendingBodyInteraction2017} demonstrated that users can maintain ownership over virtual limbs, even when these are extended or deformed. Similarly, Otono\ea\cite{otonoThirdPersonPerspectiveAvatar2022, otonoImTransformingEffects2023} found that soft-body transformations and holographic third-person views can enhance the sense of embodiment. In a more task-oriented study, Venkatakrishnan \ea\cite{venkatakrishnanGiveMeHand2023} showed that interaction performance in dense object spaces improves when users are provided with an augmented self-avatar, highlighting the value of visual hand representations in MR selection tasks.

Most existing studies focus on VR. A recent systematic review found that out of 72 papers on avatar use in XR, only six addressed MR, and just one explored asymmetric VR/MR setups \cite{weidner2023systematic}. Moreover, none of these focused on performance in selection tasks. This gap may stem in part from the lack of a consistent definition of MR across studies, as the term encompasses a wide range of technologies—from MR overlays to holographic displays and CAVEs \cite{milgram1994taxonomy}. Some MR-based Fitts’ law studies suggest that having a visible body or avatar may support performance \cite{mason2001reaching, mcgee1997fitts}, but further research is needed to evaluate these effects in more complex MR contexts

Overall, there remains a lack of direct comparative studies evaluating how different visual representations influence user experience and performance across VR and MR. This limits our understanding of the respective strengths and limitations of visual representations in each environment.

\subsection{Depth Perception}

Depth perception is critical to 3D interaction, but both VR and MR devices have several perceptual problems. The human visual system relies on a combination of monocular and binocular depth cues to estimate spatial relationships in 3D space \cite{ware2000information, hillis2004slant}. However, the accuracy of these cues can be disrupted by the technical limitations of XR technology.

In VR, one of the most significant challenges is the vergence-accommodation conflict, where, unlike in reality, the eyes converge to the depth of a perceived stimulus while accommodating (focusing) to the display surface. This leads to visual discomfort and reduces accuracy in depth-based tasks \cite{hoffman_vergenceaccommodation_2008, fernandes_looking_2025}. Several studies have shown that target depth—the distance from the user to the interaction plane— also influences pointing performance \cite{kovacs_perceptual_2008, hourcade_how_2012, kopper_human_2010}. As targets appear smaller with increasing depth, selection times tend to increase \cite{kim_effect_2025}. 

In MR/AR, depth perception is further complicated by the blending of real and virtual elements. 
Users must interpret spatial relationships between their egocentric viewpoint, physical objects, and digital overlays \cite{drascic1996perceptual, jones2008-effects, kruijff2010perceptual}. 
Studies have reported consistent depth underestimation in MR and MR \cite{jones2011-peripheral, swan2007egocentric}, although improvements have been noted in more recent systems \cite{buck2018-comparison}. 
Factors such as mismatched lighting, texture resolution, and contrast between virtual and physical content can mislead depth judgments \cite{kruijff2010perceptual, thompson2004-quality}. Occlusion errors are also a major concern: in MR, virtual content can erroneously occlude physical elements like the user’s own hands, disrupting interaction and distorting perceived spatial order~\cite{ware1995dynamic}.

Unfortunately, comparative studies about depth perception between VR and MR are scarce. To our knowledge, no rigorous evaluations have directly contrasted the effects of depth cues on user interaction across VR and MR in equivalent task settings \cite{amini_systematic_2025}. This limits our understanding of how depth perception issues manifest differently across modalities and how they may affect interaction performance.

\subsection{Fitts' Law}

\begin{figure*}
\begin{subfigure}{0.24\linewidth}
     \centering
    \includegraphics[width=\textwidth]{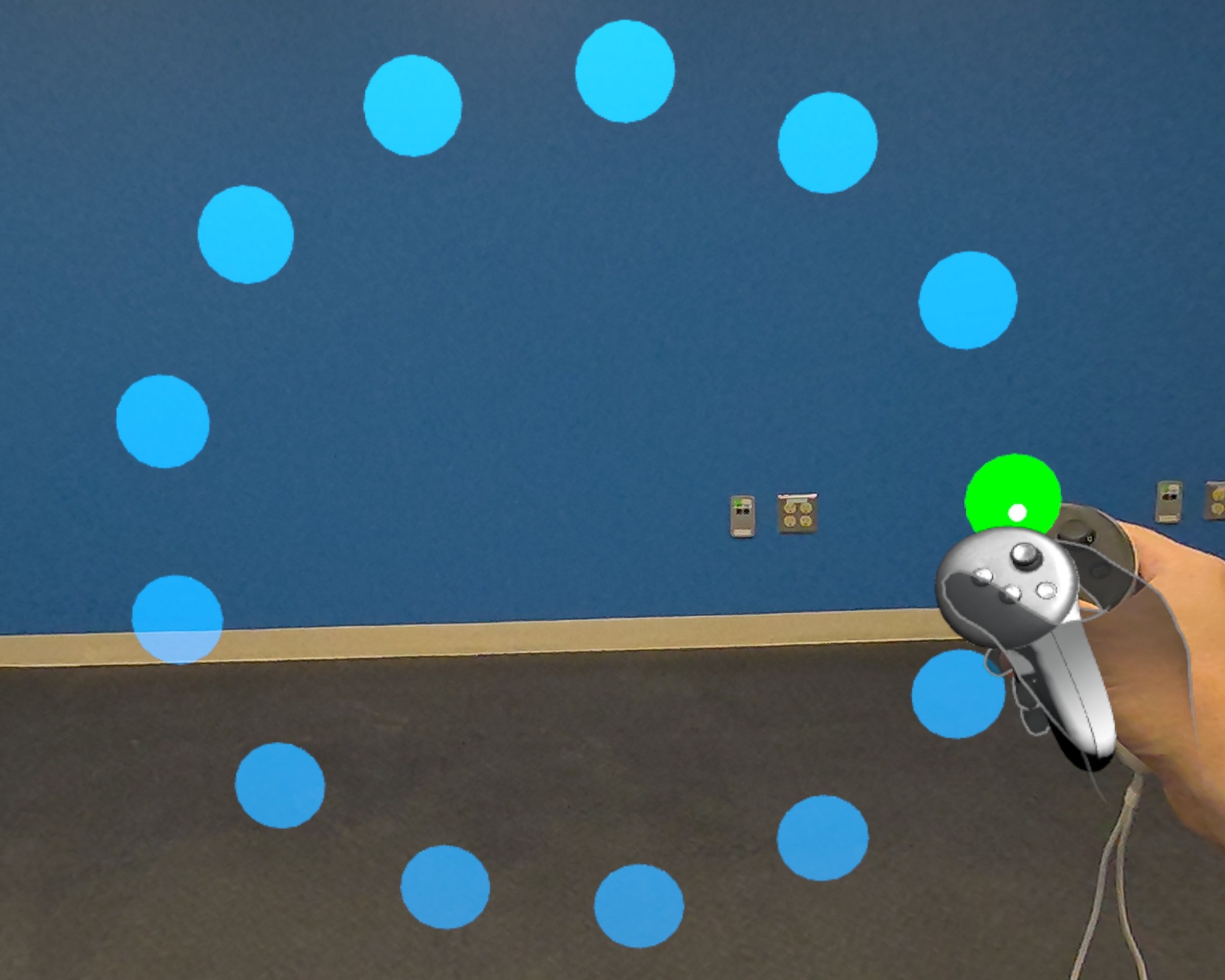}
     \caption{MR Start}
     \label{fig:fitts-example-mr-start}
\end{subfigure}
\hfill
\begin{subfigure}{0.24\linewidth}
     \centering
    \includegraphics[width=\textwidth]{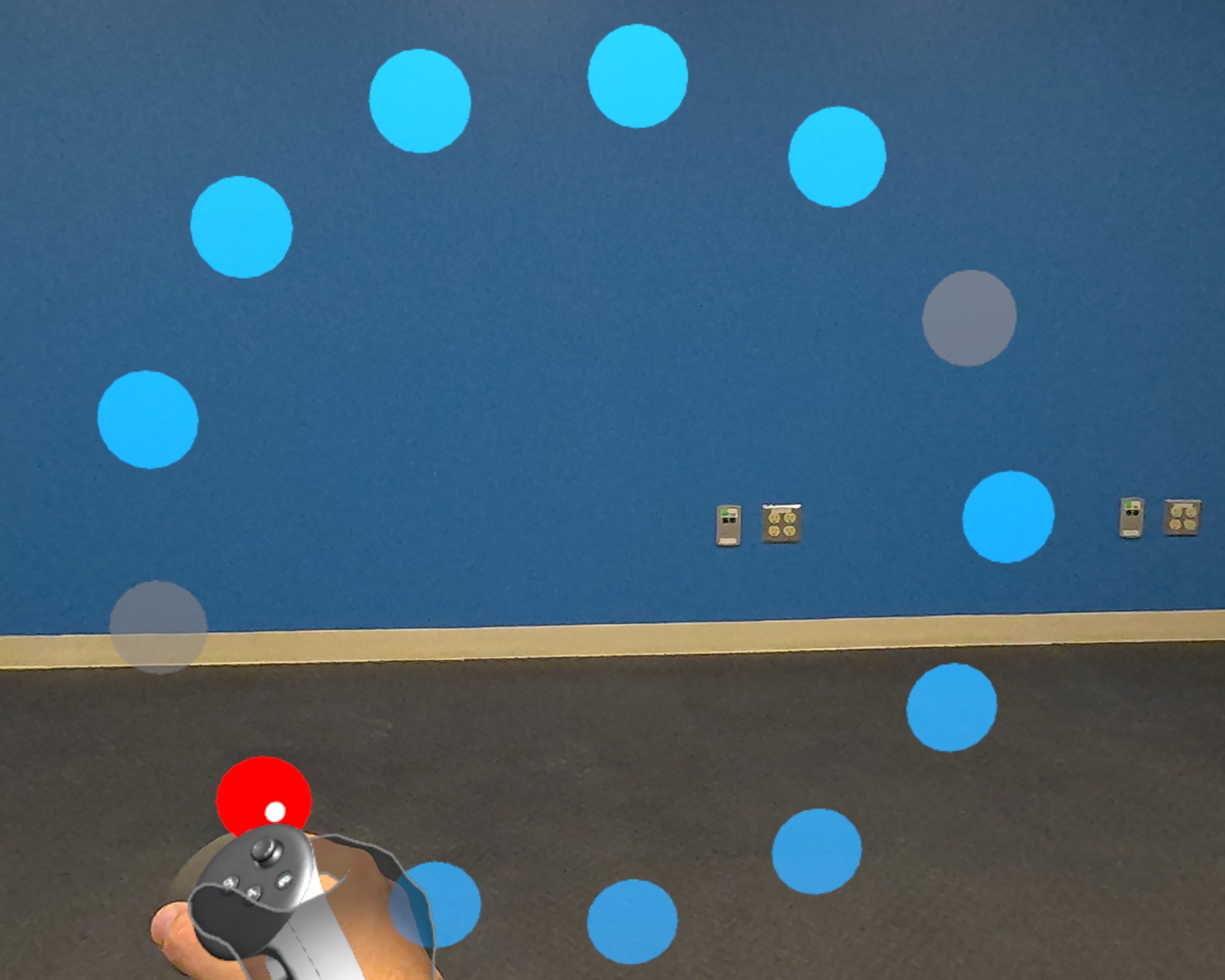}
     \caption{MR Mid}
     \label{fig:fitts-example-mr-mid}
\end{subfigure}
\hfill
\begin{subfigure}{0.24\linewidth}
     \centering
    \includegraphics[width=\textwidth]{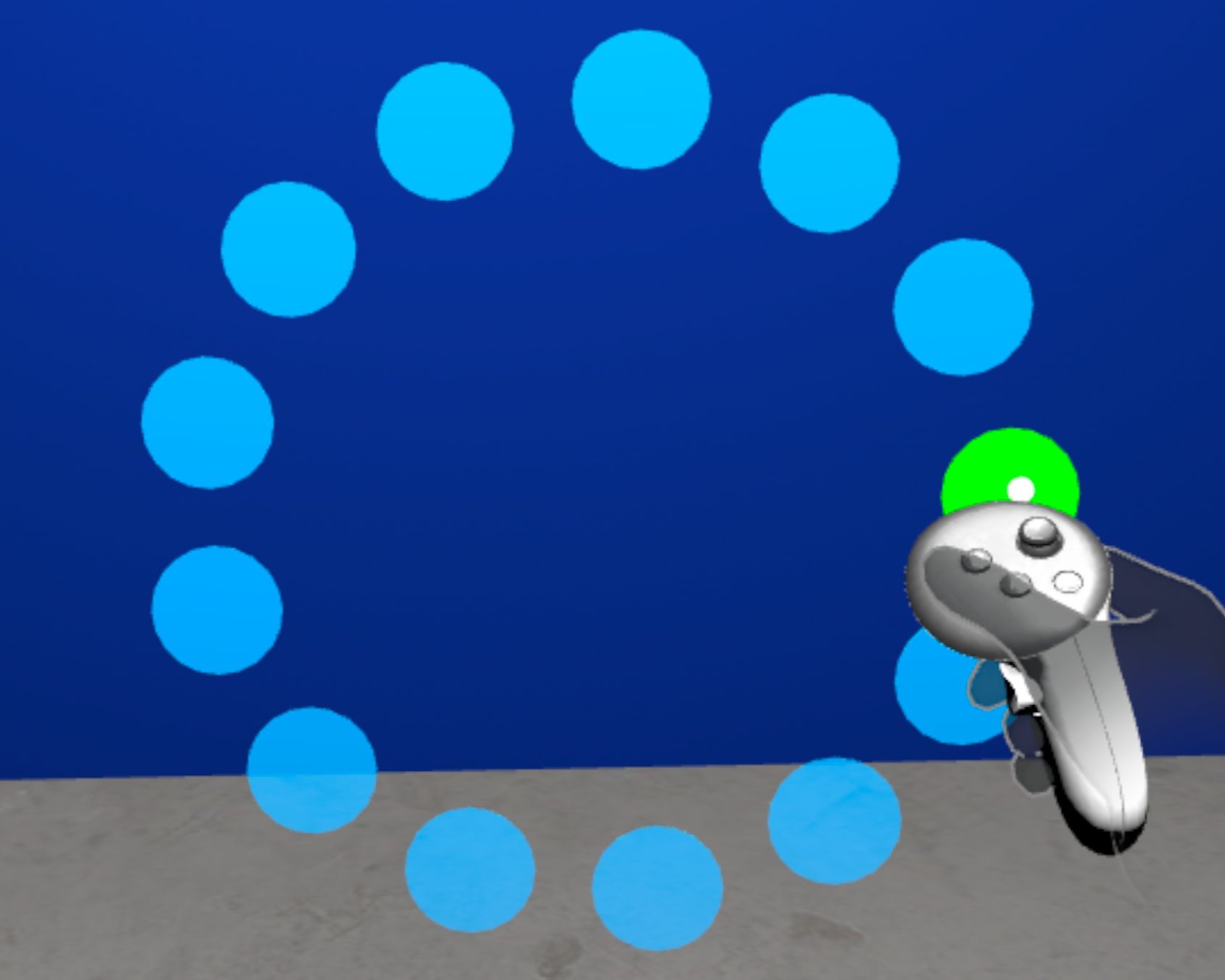}
     \caption{VR Start}
     \label{fig:fitts-example-vr-start}
\end{subfigure}
\begin{subfigure}{0.24\linewidth}
     \centering
    \includegraphics[width=\textwidth]{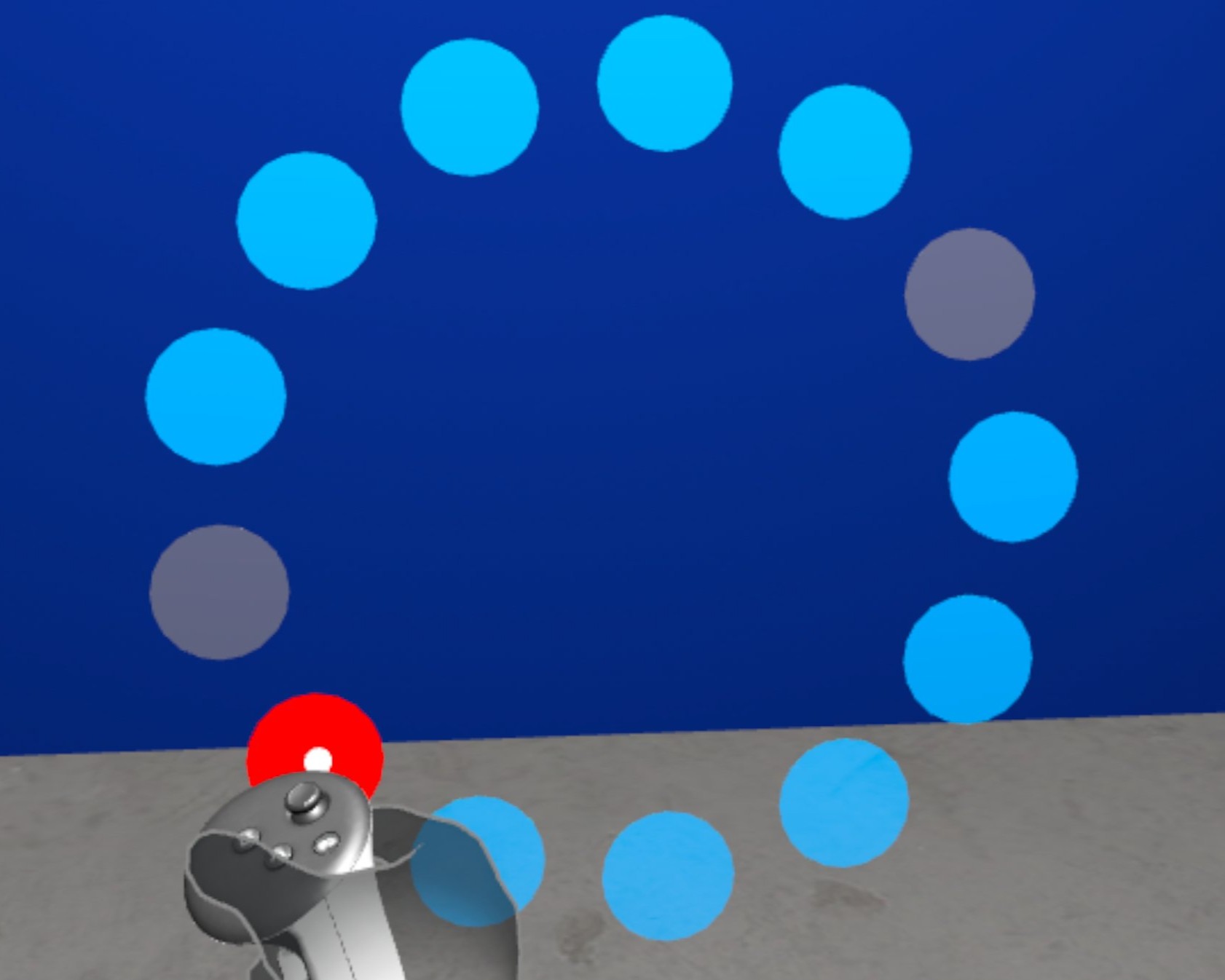}
     \caption{VR Mid}
     \label{fig:fitts-example-vr-mid}
\end{subfigure}

    \caption{The Fitts’ Law experiment, at the start and mid points of an ID, in the MR and the VR conditions. The green target indicates the participant can rest to avoid fatigue.}
    \label{fig:fitts-ex}
\end{figure*}

Since our evaluation employs the ISO 9241-411 standard \cite{iso2012human}, we describe it and its underlying model, Fitts’s law \cite{fitts1954information}, here. Fitts found that the time it takes for humans to acquire targets via rapid aimed movements depends on both the size of the target and the distance to the target from the starting position. While Fitts originally empirically validated this in a one-dimensional task, subsequent reformulations \cite{mackenzie1992extending} adapted the task to multiple directions in 2D using circular targets arranged in a ring formation. Fitts' law experiments have been effective in comparing baseline performance differences in a variety of conditions, including in XR systems \cite{amini_systematic_2025} like this work. 
Figure \ref{fig:fitts-ex} depicts a standard configuration. Typically, the task involves acquiring a target presented in a different colour from the rest in the ring; the colour indicates which circle the user should select. Movement always proceeds across the ring from one circle to the next, and the diameter of the target ring yields a consistent distance, or amplitude ($A$), for each selection. Target width ($W$) is also varied to produce different task difficulties, presented as the index of difficulty ($ID$) measured in bits, see equation (1): 

\begin{equation}
    ID = \log_2 \left(\frac{A}{W} + 1\right)
\end{equation}

Fitts' law models the relationship between ID and recorded mean movement time ($MT$) as a linear equation, as seen in equation 2:
\begin{equation}
    MT = a + b \times ID
\end{equation}

where a and b are derived via linear regression.
Overall selection efficiency is determined using $ID$ and $MT$, yielding throughput ($TP$) measured in bits per second. See equation (3):

\begin{equation}
    TP = \frac{ID}{MT}
\end{equation}

This metric facilitates comparison of the overall effectiveness of an input device or conditions in which it was used to known baselines, supporting the design of new interfaces \cite{macKenzie1992fitts}. 

 Throughput can be further enhanced through the use of so-called “effective” measures, which account for the task participants actually perform in the study rather than that presented. Effective width ($W_e$) replaces the presented target width ($W$) and is calculated through equation (4):

\begin{equation}
    W_e = 4.133 \times SD_x
\end{equation}

$SD_x$ is the standard deviation of selection coordinates along the task axis. $W_e$ yields an effective target size where 96\% (i.e., $\pm$ 2.066 standard deviations from the mean) of selections would have hit the target, adjusting experimental accuracy and facilitating TP comparison between studies with different error rates. Effective amplitude ($A_e$) is the average of the actual movement distance rather than the presented target distance. Equation (5) includes effective width and amplitude, yielding the effective index of difficulty ($ID_e$), in bits, adjusting for accuracy:

\begin{equation}
    ID_e = \log_2 \left(\frac{A_e}{W_e} + 1\right)
\end{equation}

Effective throughput, $TP_e$, employs $ID_e$ in the same fashion that standard throughput uses $ID$. See equation (6):

\begin{equation}
    TP_e = \left(\frac{ID_e}{MT}\right)
\end{equation}

The main advantage of effective measures is their consistency in throughput calculations. Effective throughput is less susceptible to speed/accuracy tradeoffs \cite{mackenzie2008fitts}. This consistency is important in cross-study comparisons of devices. For example, from extensive prior research employing the standard, the consensus is that the mouse is among the most performant input devices, with a throughput score of around 4.5 bps \cite{heidicker2017influence, knierim2018physical}. Fitts’ law has been used in XR systems to test different input devices and the impacts of variations in VEs on the ability of users to interact with the system effectively \cite{mcgee1997fitts} and also to examine haptic feedback's role in VR selection tasks \cite{kourtesis2022action}.

\section{Methodology} \label{methodology}

In this section, we present the methodology of the user study we conducted using Fitts' law to evaluate the impact of controller representation in MR. 

\subsection{Participants}  \label{participants}

We recruited 40 participants for our study. This included 20 women, 17 men, 1 gender fluid individual, 1 transgender woman, and 1 who declined to answer; the mean age was 23.95 years ($SD$ = 6.06). All participants self-reported normal stereoscopic vision and were able to operate one Meta Quest Touch controller. A total of 37 participants reported being right-handed, 2 left-handed, and one reported being ambidextrous but used their right hand for the study. Participants were asked to self-report how frequently they played video games, used virtual reality (VR), and participated in activities that require a high degree of hand-eye coordination (e.g., sports), summarized in Table \ref{tab:participant-gaming}. All participants were compensated with \$15 in local currency for their participation. Our experimental protocol was subject to and passed ethical review by our institution's ethics board.

\begin{table}[h]
    \caption{Self-reported Likert scale frequency of video gaming, using VR, and participating in activities requiring a high degree of hand-eye coordination. Never = 1, Regularly = 5.}
    \centering
    \begin{tabular}{lccc}
        \toprule
        Frequency & Video Game & VR & Hand-eye coord. activities \\
        \midrule
        Never     & 11 & 5  & 7  \\
                  & 8  & 3  & 13 \\
                  & 14 & 7  & 7  \\
                  & 6  & 7  & 6  \\
        Regularly & 1  & 18 & 7  \\
        \bottomrule
    \end{tabular}
    \label{tab:participant-gaming}
\end{table} 

\subsection{Apparatus}  \label{apparatus}

\subsubsection{Hardware}  \label{hardware}
We used a Meta Quest 3 VR headset with a 2064 x 2208 pixels per eye resolution, a native 90 Hz refresh rate, and an effective field of view (FOV) of 110° horizontal and 96° vertical. The HMD features an internal processor and storage, allowing it to function without cables, eliminating movement restrictions. The HMD uses external-facing, high-resolution, full-colour cameras with infrared depth sensing for tracking. These cameras also support an MR pass-through mode, which we used for the MR conditions in our experiment. The HMD comes with two tracked hand-held controllers, which we used for the selection task and other interactions in the VEs.

\subsubsection{Software}  \label{software}

We developed the experiment software in Unity 3D (version 6000.0.37f1) using the Meta All-In-One Software Development Toolkit (SDK) (version 72.0.0) for headset and controller tracking, interactions, and access to pass-through cameras. The software presented an extended reality implementation of the standardized ISO-9241-411 task \cite{iso2012human}. The application displayed a ring of circular targets in front of participants for selection, as seen in Figure \ref{fig:fitts-ex}. The software presented targets as flat circles to restrict the depth selection plane to the width of the targets. The system supported any combination of target count within a ring, and any combination of amplitudes and widths, as required by the experiment. The system distributed the experiment conditions and the Fitts' law \emph{ID} configurations described in Section \ref{design} in balanced Latin squares, selected on the participant number by a modulus calculation. For example, participant number 14 would have condition group 6 (14 mod 8) and Fitts' law group 2 (14 mod 12). The targets changed colour depending on their state: the starting target before the timer started was green (see Figure \ref{fig:fitts-example-mr-start} allowing the participant to rest before beginning if they felt fatigued \ref{fig:fitts-example-vr-start}). Otherwise, the active target to be selected was red , previously selected targets were gray, and the remaining unselected targets were blue (see Figure \ref{fig:fitts-example-mr-mid} and \ref{fig:fitts-example-vr-mid}). The software also supported having targets optionally turn orange when the poke interactor made contact with them. This feature was enabled only in practice sessions to demonstrate how to make contact with a target. It was disabled during actual experiment sessions, so participants had to judge the depth accuracy of their selections without the extra feedback of the colour change on contact.

\begin{figure}[h]
    \centering
    \begin{subfigure}[b]{0.49\columnwidth}
        \centering
        \includegraphics[width=\textwidth]{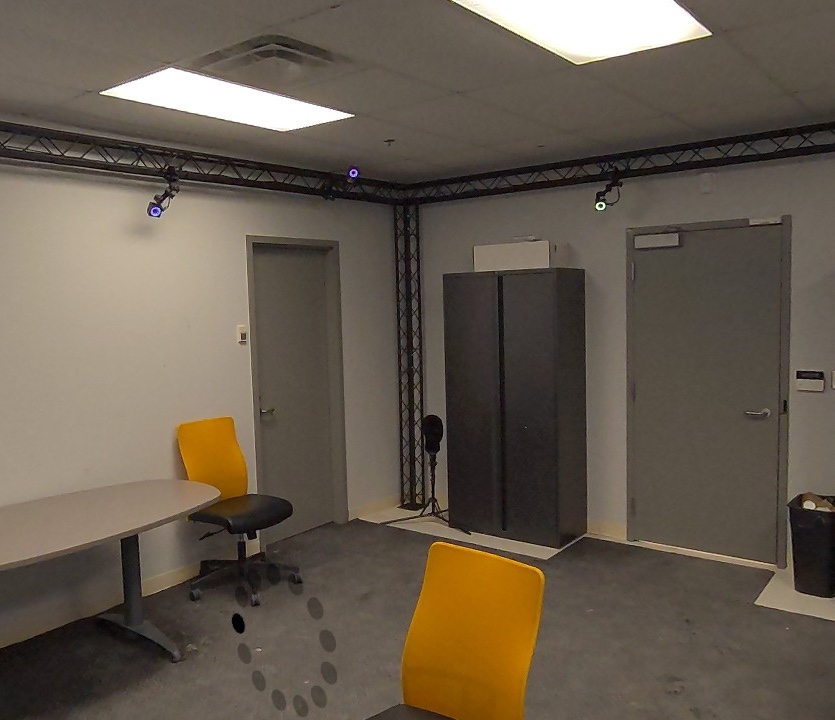}
        \vfill
        \caption{MR Room}
        \label{fig:mr-room}
    \end{subfigure}
    \hfill
    \begin{subfigure}[b]{0.49\columnwidth}
        \centering
        \includegraphics[width=\textwidth]{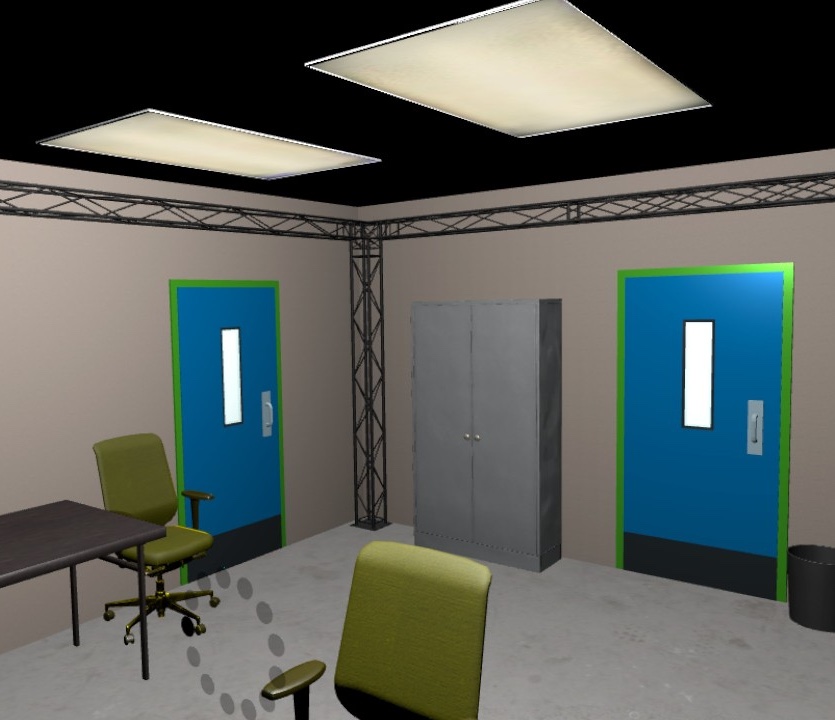}
        \caption{VR Room}
        \label{fig:vr-room}
    \end{subfigure}
    \hfill
    \caption{The real-world experiment location visible in MR and the VR recreation of the location.}
    \label{fig:room-images}
\end{figure}

Participants selected targets using direct touch via a "poke" interaction. This used the SDK's default poke interactor object, which is the small sphere visible in Figure ~\ref{fig:conditions-no-task}b. The sphere follows the tracked controller and provides information on collisions with objects in the virtual environment. This interactor was 1 cm in diameter in the virtual environment, which effectively increases the diameter of the targets by 1 cm. The system used the index trigger on the controller to confirm selection. On selection, if the poke interactor intersected the target, the selection was recorded as a "hit". Otherwise, the selection was considered a "miss" to facilitate the calculation of the error rate. The software played a sound upon pressing the trigger, regardless if the selection was a hit or a miss. Each time the trigger was pressed, the software recorded the selection time, the controller position (captured as a three-dimensional vector), whether the selection hit or missed, as well as the target's amplitude and width. All data was automatically logged in a participant ID and date, time-stamped comma-separated value file. If any unselected targets remained, the next target in the sequence (i.e., the one directly across the ring from the current one, following the ISO standard task) would then activate.

Depending on the condition, the selection task was displayed in either a "VR room" or "MR room", see Figure \ref{fig:room-images}. We referred to this independent variable as XR Mode. The VR environment featured a modestly detailed virtual replica of the room where the study took place, which accurately approximated the exact dimensions and layout of the furniture in the study location (see Figure \ref{fig:room-images}). The second was an MR environment, where the only virtual content displayed was the virtual selection targets and the controller representation. In the MR environment, there were no other visible virtual objects. The Quest 3 pass-through cameras were activated in this condition, allowing participants to see the real-world room. 

We refer to the controller representation independent variable as "Controller Mode". Depending on the Controller Mode condition, the software would change the representation of the tracked controller in the virtual scene, as follows:

\begin{itemize}
    \item \textbf{Controller \textendash} The virtual controller representation was a white Meta Quest 3 controller model with a dark grey face (Figures \ref{fig:conditions-no-task}a and e).
    \item \textbf{Hand \textendash} The virtual hand representation was a dark grey semi-transparent virtual hand model with a light grey outline (Figures \ref{fig:conditions-no-task}b and f).
    \item \textbf{Both \textendash} Both the virtual controller representation and virtual hand representation were displayed (Figures \ref{fig:conditions-no-task}c and g).
    \item \textbf{No Representation (None) \textendash} Neither the hand nor the controller were visible. Figures \ref{fig:conditions-no-task}d and h).
\end{itemize}

Note that regardless of the Controller Mode, the Poke Interactor sphere was always visible.

\subsubsection{Experimental Task}  \label{task}

Our experimental task followed the ISO 9241-411 standard \cite{iso2012human}. The software was configured to present rings of 13 targets placed at three different amplitudes (distances): 30 cm, 45 cm, and 60 cm, and four widths: 5 cm, 15 cm, 20 cm, and 25 cm. This yielded 12 $ID$s (see equation 1), providing a range from 1.0 to 3.5, which is within the common range of IDs for XR Fitts' law studies \cite{amini_systematic_2025}, without the amplitude exceeding a comfortable distance for a participant to reach from a seated position. The task involved selecting the highlighted target by reaching out using the controller, touching the spherical cursor at its tip against the active target directly, and pressing the trigger button to select a target. When all 12 $ID$s for a condition were completed, a UI button appeared, which the participant selected to proceed to the next condition. Participants could take breaks on this next condition screen, or after each set of 13 targets, i.e., before selecting the first target in the sequence, to prevent participant fatigue.

\subsection{Procedure}  \label{procedure}

We welcomed participants to the experiment space and asked each to complete a consent form and a demographic survey. We then explained the task of the experiment to the participants. Participants remained seated throughout the study. After participants put on and adjusted the Quest 3, they were presented with a user interface that required them to enter their participant number (to handle condition ordering) and their preferred dominant hand for the entire study. Participants adjusted the position (height and distance) of the targets to reach them all comfortably through a calibration process. Then they completed a practice session for the selection task with 1 block of discs where amplitude = 30 cm and width = 5 cm, and 1 block where amplitude = 60 cm and width = 25 cm. In these practice trials only, the target was highlighted in orange upon intersecting the poke interactor with the target; this feedback was intended to show participants what a successful selection would look like. After completing the practice trials, participants continued to the actual study. Participants were instructed to select the highlighted target as quickly and as accurately as possible. Upon completion of all trials, participants filled out a post-study questionnaire. The questionnaires were custom-designed and asked how \emph{noticeable} the changes in Controller Mode were in VR and MR as a 5-point Likert scale question. This was followed by questions for which Controller Mode was \emph{preferred} in both VR and MR, and which XR Mode was \emph{preferred}. Finally, the participant was asked to provide any additional comments explaining their choices for each of these questions. We created these questionnaires to evaluate participants' subjective perceptions of both the XR and controller conditions, allowing for comparisons with our dependent variables and providing further insights. The participant was then compensated for their participation, and the HMD was then sanitized using a CleanBox and reset for the next participant. The sessions were scheduled for 60 minutes and took, on average, 44 minutes.

\subsection{Experimental Design}  \label{design}

Our experiment employed a 2 $\times$ 4 within-subjects design with the factors XR Mode (VR, MR) and Controller Mode (Controller, Hand, Both, and None), as detailed above. We counterbalanced the ordering of the eight combinations of XR Mode and Controller Mode according to an 8 x 8 balanced Latin square. We included 3 target amplitudes (30, 45, 60 cm) and 4 target widths (5, 15, 20, 25 cm) to generate a range of $ID$s per Fitts’ law. As we regarded a correct selection as intersecting the 1 cm poke interaction (see Section \ref{software}) with the target, in our Fitts' law analysis and calculation of error rate, widths were 6 cm, 16 cm, 21 cm, and 26 cm \cite{kabash_prince_1995}. These yielded 12 unique IDs ranging from 1.11 to 3.46 bits, distributed using a 12 $\times$ 12 Latin square. In summary, participants completed: 2 XR Modes $\times$\ 4 Controller Modes $\times$\ 12 IDs $\times$\ 13 Selection Trials = 49,920 selections total (or 1,248 selections per participant).

The dependent variables included target selection time (in ms), error rate (\% of missed selections), effective throughput (in bps, calculated using Equation 6), and depth deviation (in cm). Depth deviation was calculated as the shortest distance between the poke interactor and the target's z coordinate, set to zero when the poke interactor intersects the target. Subjective feedback is reported from the survey described in Section \ref{procedure}.
\section{Results}

We conducted data analyses on aggregate data (means) grouped by Controller Mode, XR Mode, $ID_e$ for Throughput (see Equation 6), and ID for Linear Regression Plots (see Equation 1). We conducted Shapiro-Wilk tests of normality on each of our dependent variables and found that our data violated the assumption of normality. Therefore, we performed an aligned-rank transform \cite{art2011} for each metric, followed by two-way ANOVAs. Post-hoc comparisons were conducted using ART-C \cite{elkin2021-artc}.

\begin{figure*}
\begin{subfigure}{0.32\linewidth}
    \small
    \begin{tabular}[b]{llccc}
    \toprule
    XR & Controller & Intercept & Slope & $R^2$ \\
    \midrule
    VR & None        & 0.344 & 0.148 & 0.929 \\
    VR & Hand        & 0.360 & 0.123 & 0.849 \\
    VR & Controller  & 0.388 & 0.107 & 0.805 \\
    VR & Both        & 0.365 & 0.110 & 0.829 \\
    \midrule
    MR & None        & 0.346 & 0.147 & 0.905 \\
    MR & Hand        & 0.358 & 0.118 & 0.805 \\
    MR & Controller  & 0.372 & 0.116 & 0.787 \\
    MR & Both        & 0.353 & 0.113 & 0.791 \\
    \bottomrule
    \\
    \end{tabular}
    \caption{Linear regressions across XR and controller modes.}
    \label{fig:fitts-tab}
\end{subfigure}
\hfill
    \begin{subfigure}[b]{0.3\textwidth}
        \centering
        \includegraphics[width=\linewidth]{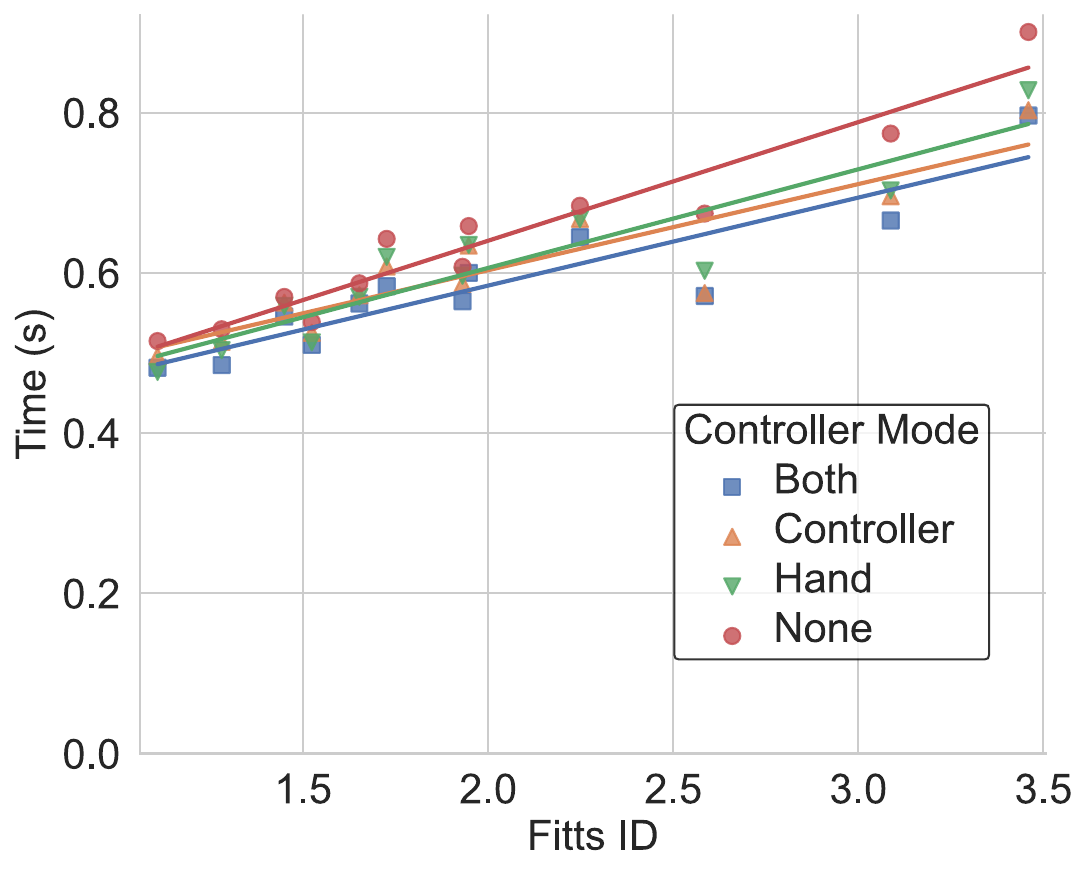}
        \caption{Linear regression plots in VR by controller modes.}
        \label{fig:fitts-vr}
    \end{subfigure}
    \hfill
    \begin{subfigure}[b]{0.3\textwidth}
        \centering
        \includegraphics[width=\linewidth]{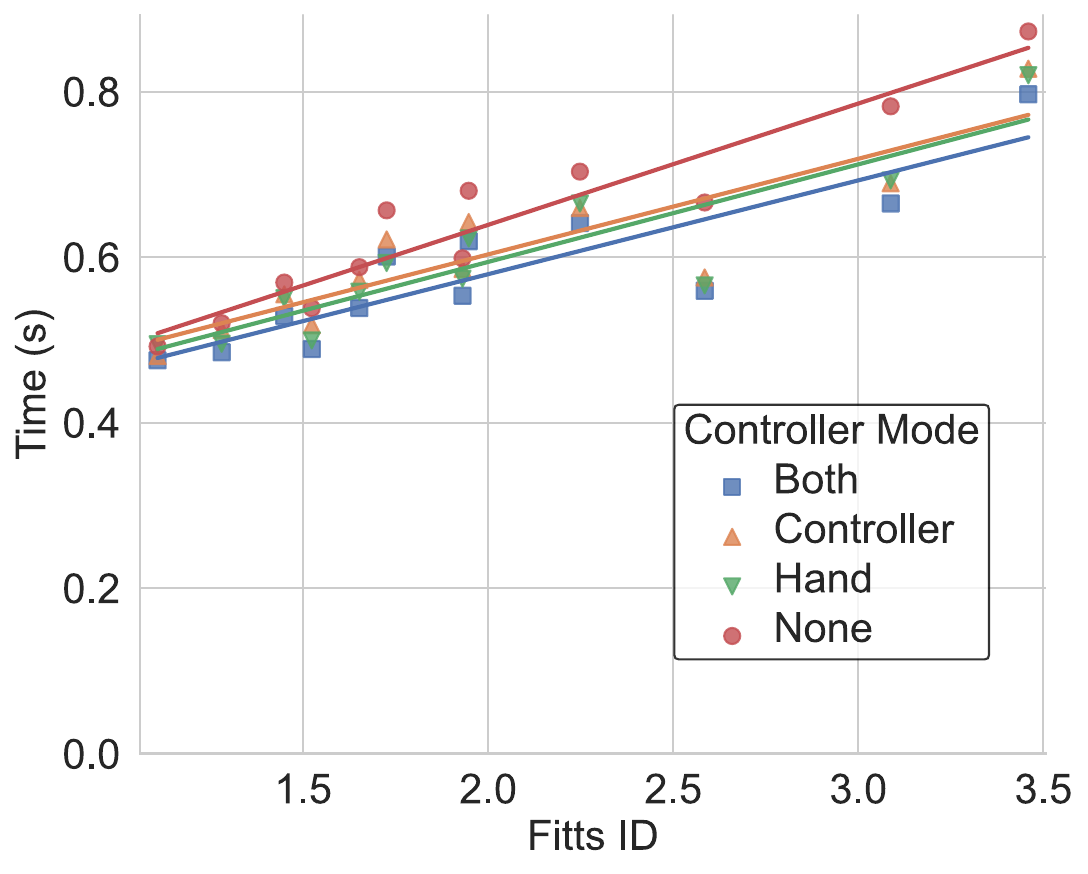}
        \caption{Linear regression plots in MR by controller modes.}
        \label{fig:fitts-mr}
    \end{subfigure}
    \caption{Fitts' law analysis by XR mode and controller modes.}
    \label{fig:fitts-all}
\end{figure*}

\subsection{Selection Time}

We found a significant main effect of Controller Mode on movement time ($F_{3, 117} = 8.69$, $p < .001$, $\eta_p^2 = 0.18$). See Figure~\ref{fig:mt}. The None condition had the highest movement time ($M = 0.640$ s, $SD = 0.300$), followed by Controller ($M = 0.603$ s, $SD = 0.268$), Hand ($M = 0.600$ s, $SD = 0.273$), and Both ($M = 0.582$ s, $SD = 0.238$). Post-hoc comparisons revealed significantly higher movement time in the None condition compared to Both ($p < .001$), Controller ($p = .008$), and Hand ($p < .0001$). There were no significant effects of XR Mode ($F_{1, 39} = 0.661$, $p = 0.421$), or significant interactions ($F_{3, 117} = 0.624$, $p = 0.601$). See Figure \ref{fig:mt}. Mean selection time across XR conditions was: MR ($M = 0.604$ s, $SD = 0.277$) and VR ($M = 0.608$ s, $SD = 0.267$).

As is commonly done in studies conforming to ISO 9241-9 methodology, we also derived regression models of $MT$ in terms of $ID$. The linear regression parameters of these Fitts' law models are seen in Figure \ref{fig:fitts-tab}, separated by both Controller Mode and XR Mode. These are graphically depicted in Figure \ref{fig:fitts-vr} (VR mode) and Figure \ref{fig:fitts-mr} (MR mode).

\subsection{Effective Throughput}

We found a significant main effect of Controller Mode on effective throughput ($F_{3, 117} = 16.33$, $p < .001$, $\eta_p^2 = 0.30$). See Figure~\ref{fig:tpe}. The None condition had the lowest effective throughput ($M = 5.00$ bps, $SD = 1.55$), compared to Controller ($M = 5.37$ bps, $SD = 1.60$), Hand ($M = 5.47$ bps, $SD = 1.67$), and Both ($M = 5.46$ bps, $SD = 1.55$). Post-hoc comparisons revealed significantly lower effective throughput in the None condition compared to Both, Controller, and Hand (all $p < .0001$). There were no significant effects of XR Mode ($F_{1, 39} = 0.341$, $p = 0.563$), or significant interactions ($F_{3, 117} = 0.065$, $p = 0.978$). See Figure~\ref{fig:tpe}. Effective throughput across XR conditions was: MR ($M = 5.355$ bps, $SD = 1.624$) and VR ($M = 5.294$ bps, $SD = 1.585$).

\subsection{Depth Deviation}

\begin{figure*}
\begin{subfigure}{0.32\linewidth}
    \includegraphics[width=\columnwidth]{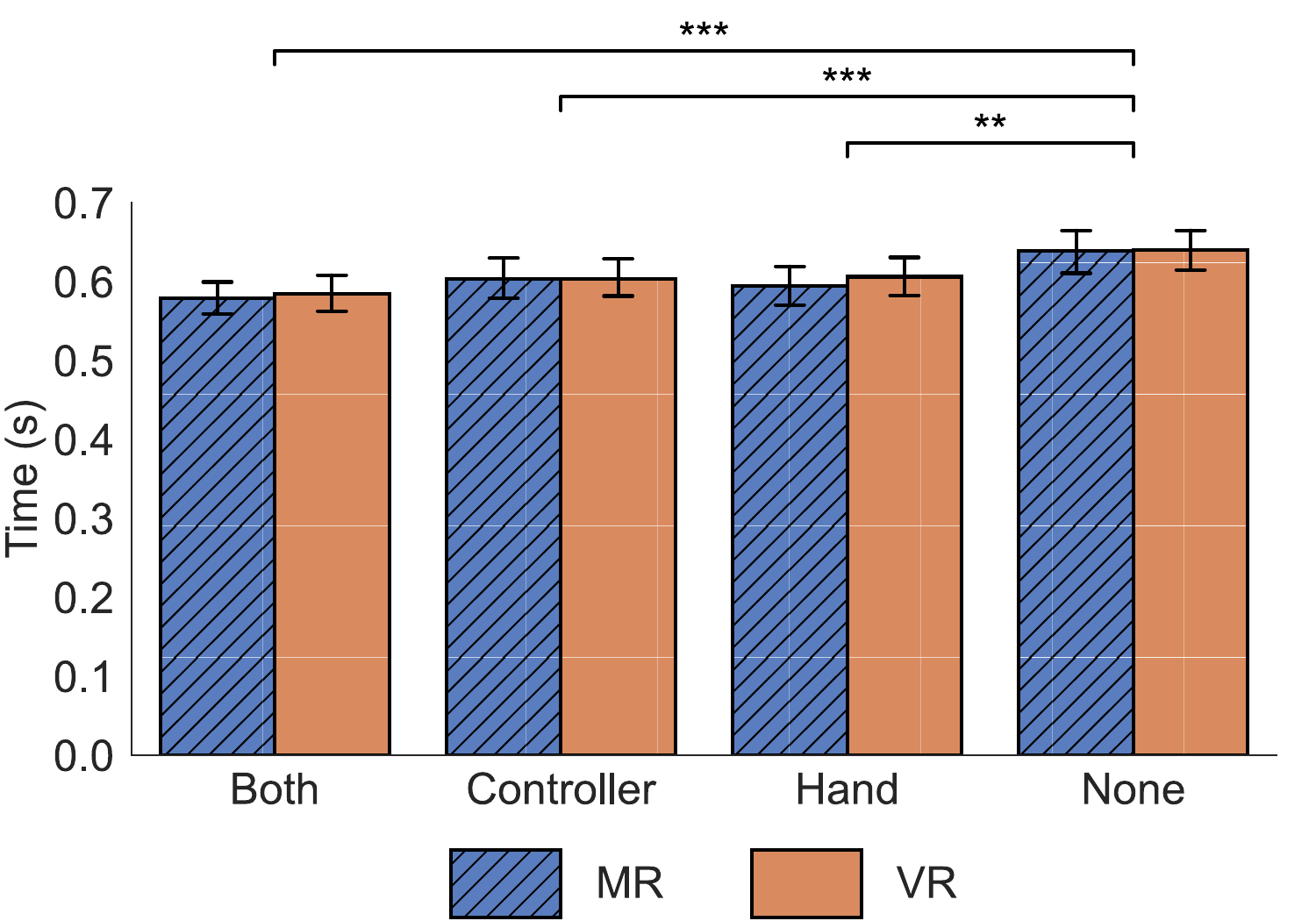}
    \caption{Mean selection time in seconds across controller and XR modes. }
    \label{fig:mt}
\end{subfigure}
\hfill
\begin{subfigure}{0.32\linewidth}
  \centering
    \includegraphics[width=\columnwidth]{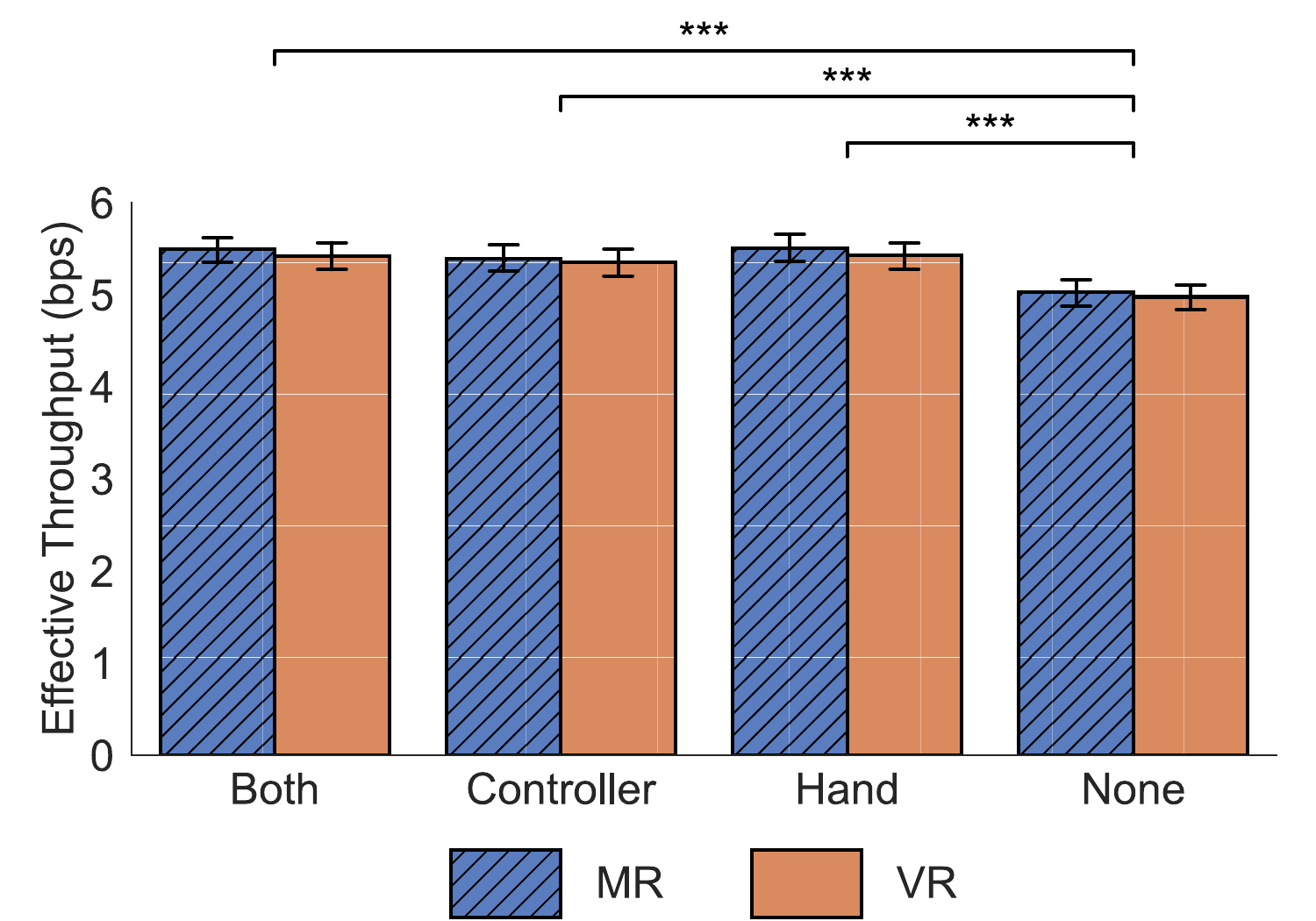}
    \caption{Effective throughput across controller and XR modes.}
    \label{fig:tpe}
\end{subfigure}
\hfill
\begin{subfigure}{0.32\linewidth}
  \centering
    \includegraphics[width=\columnwidth]{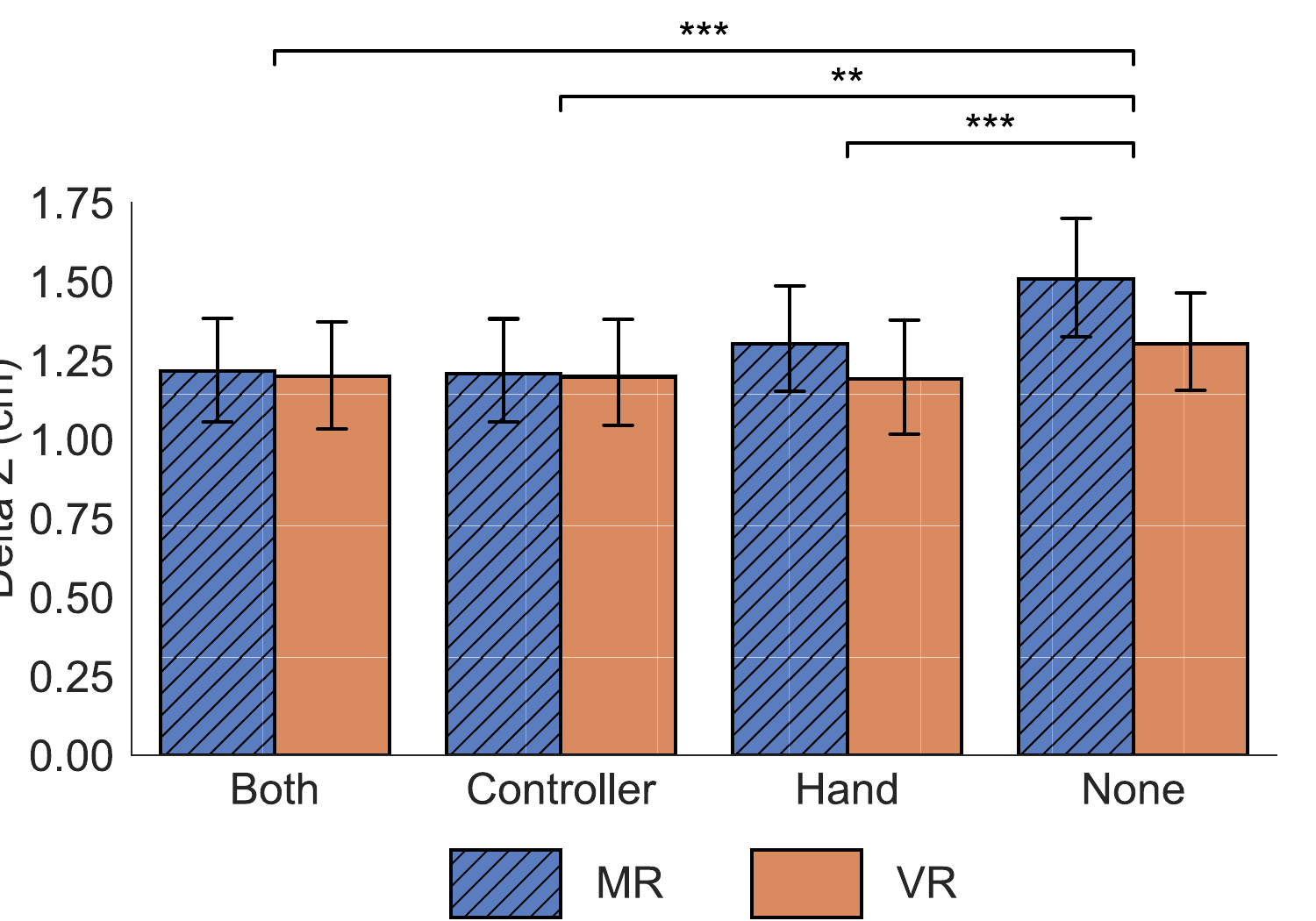}
    \caption{Depth deviation (Delta Z) in cm across controller and XR modes. }
    \label{fig:depth}
\end{subfigure}

    \caption{Selection Time, effective throughput, and depth deviation (Delta Z) across controller and XR mode. Error bars indicate a 95\% confidence interval. Significance is indicated at each level: $<$ 0.001 = ***, $<$ 0.01 = **, $<$ 0.05 = *.}
    \label{fig:fitts-all}
\end{figure*}

We found a significant main effect of Controller Mode on depth deviation 
($F_{3, 117} = 7.24$, $p < .001$, $\eta_p^2 = 0.16$). See Figure~\ref{fig:depth}. The None condition resulted in the highest depth deviation ($M = 1.41$ cm, $SD = 1.97$), compared to Hand ($M = 1.25$ cm, $SD = 2.02$), Both ($M = 1.21$ cm, $SD = 1.98$), and Controller ($M = 1.20$ cm, $SD = 1.80$). Post-hoc comparisons revealed significantly greater depth deviation in the None condition compared to Both ($p < .001$), Controller ($p = .005$), and Hand ($p < .001$). There were no significant effects of XR Mode ($F_{1, 39} = 2.25$, $p = 0.142$), or significant interactions ($F_{3, 117} = 1.91$, $p = 0.132$). See Figure~\ref{fig:depth}. Depth deviation across XR conditions was: MR ($M = 0.0131$ cm, $SD = 0.0197$) and VR ($M = 0.0122$ cm, $SD = 0.0192$).

 \begin{figure}[hpb!]
  \centering
    \includegraphics[width=\columnwidth]{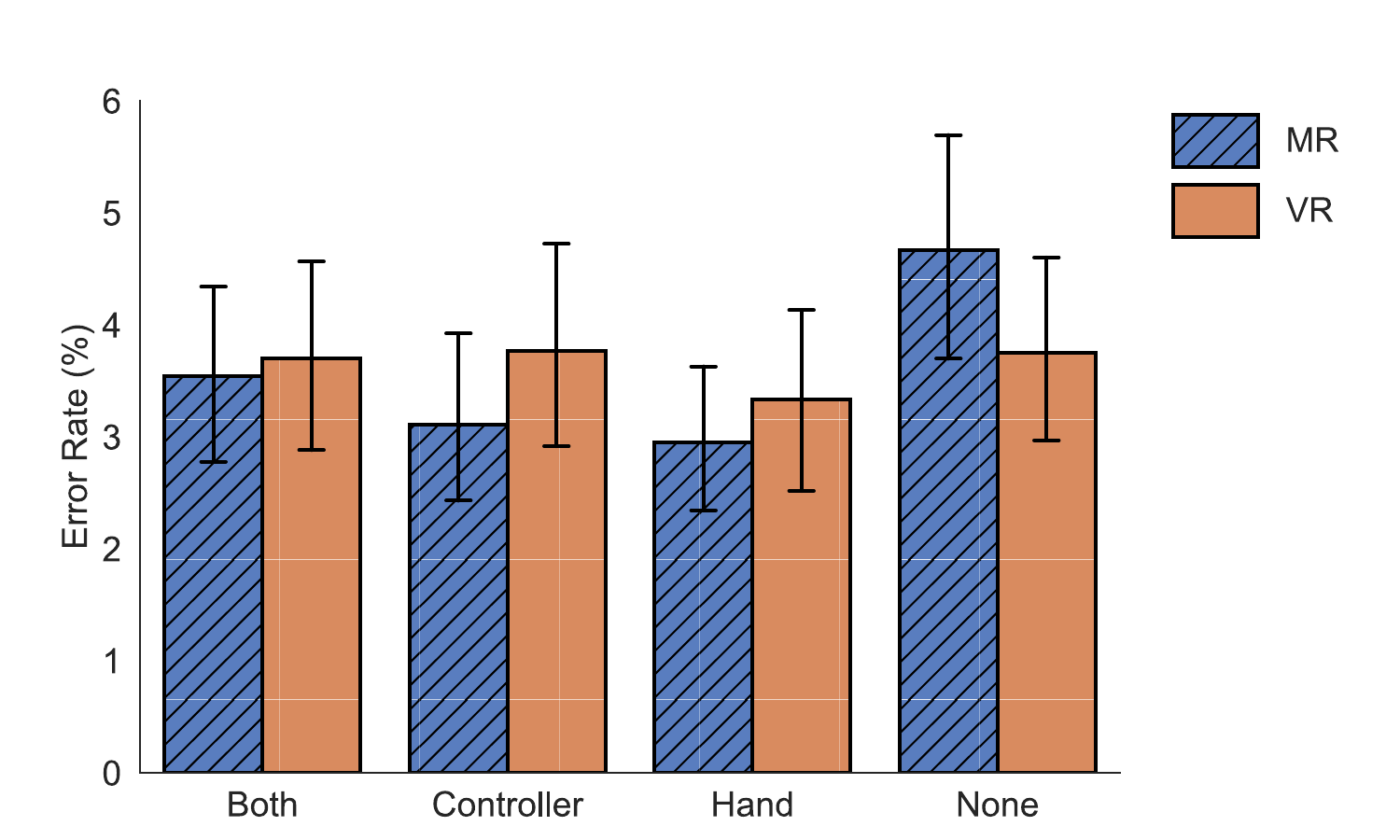}
    \caption{Error rate across controller and XR modes. Error bars indicate a 95\% confidence interval.}
    \label{fig:error}
\end{figure}

\subsection{Error Rate}

We calculated the error rate as the deviation on the XY plane beyond the edge of the target. Due to a ceiling effect in the error rate data, where most responses were correct, there were insufficient error observations to support a two-factor analysis. To explore potential differences by condition, we conducted separate one-way ANOVAs on the aligned rank-transformed data for XR Mode and Controller Mode individually. There were no significant differences for either XR Mode ($F_{1, 39} = 0.019$, $p = 0.891$) or Controller Mode ($F_{3, 117} = 2.47$, $p = 0.065$). See Figure~\ref{fig:error}. For XR Mode, mean error rates were: MR ($M = 3.57$\%, $SD = 9.08$\%) and VR ($M = 3.64$\%, $SD = 9.18$\%). Mean error rates by Controller Mode were: Both ($M = 3.62$\%, $SD = 9.01$\%), Controller ($M = 3.44$\%, $SD = 9.02$\%), Hand ($M = 3.14$\%, $SD = 8.05$\%), and None ($M = 4.21$\%, $SD = 10.29$\%).

\subsection{Subjective Results}

\subsubsection{Noticeability of changes in controller modes}

\begin{figure}[hb!]
    \centering
    \includegraphics[width=\columnwidth]{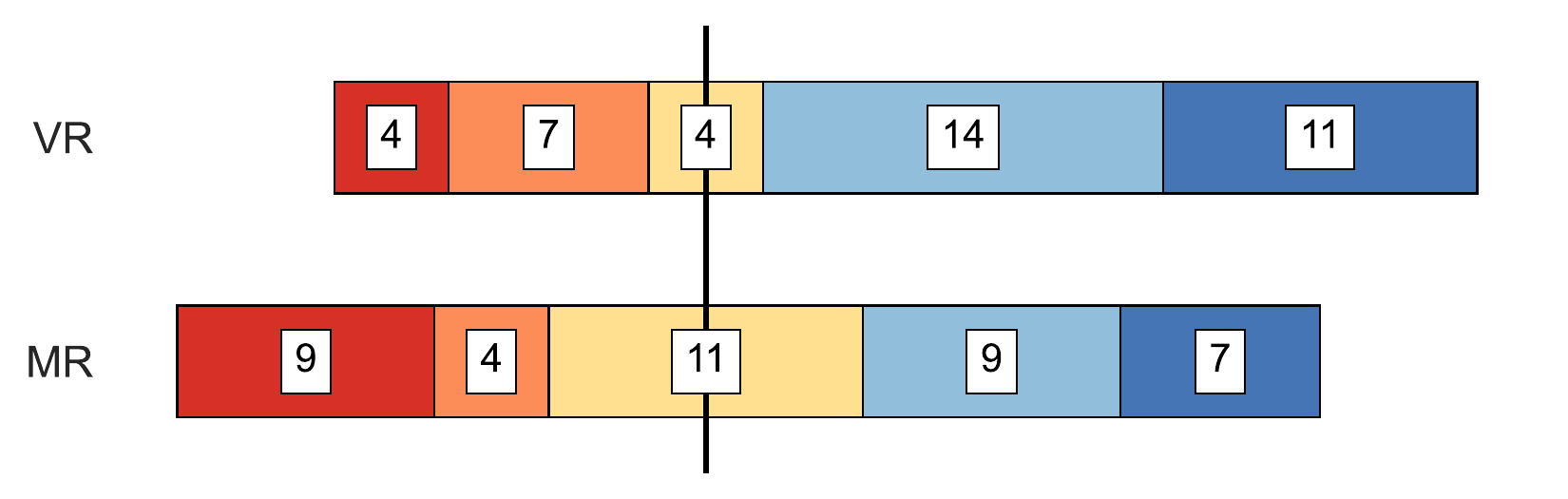}
    \caption{Noticeability ratings of changes in controller modes across XR modes; 1 = Not noticeable, 5 = Very noticeable. Bar labels indicate the number of participants who selected each rating.}
    \label{fig:xr-notice}
\end{figure}

From our Noticeability Likert scale metric, an aligned-rank transformed one-way ANOVA revealed changes in controller modes in VR were significantly more noticeable than those in MR ($F_{1, 39} = 6.01$, $p = .019$, $\eta^2 = .13$). see Figure~\ref{fig:xr-notice}.

\subsubsection{Preference and Insights}\label{sec:preference}

Participants completed a post-study preference questionnaire to gather subjective opinions on the conditions. In terms of XR Mode, of the 40 participants, 18 preferred VR conditions, 14 preferred MR conditions, and 8 had no preference. 
Most participants who preferred the VR mode noted that they felt it was less distracting, as the real-world lab contained additional objects and details that were not visible in the VR mode. Some indicated that they enjoyed the novelty of having the entire environment replaced, rather than seeing the real world. Several participants also noted a visible "lag" or "delay" in the MR conditions, with the controller representations moving slightly behind their physical hand and controller. P27 stated, "The image of the hand and the controller didn't match the real world hand and controller exactly, there was a small amount of lag, and I didn't like seeing a double of them." Participants who preferred the MR mode indicated that they felt more comfortable knowing what was happening around them, compared to the VR mode, where they could not see what was happening in the room. They also reported that seeing their arms made the selection process more realistic and easier, as P36 indicated, "I was more used to those specific conditions where I could gauge things around me better (such as seeing my arms)."

\begin{figure} [hb]
    \centering
    \includegraphics[width=\columnwidth]{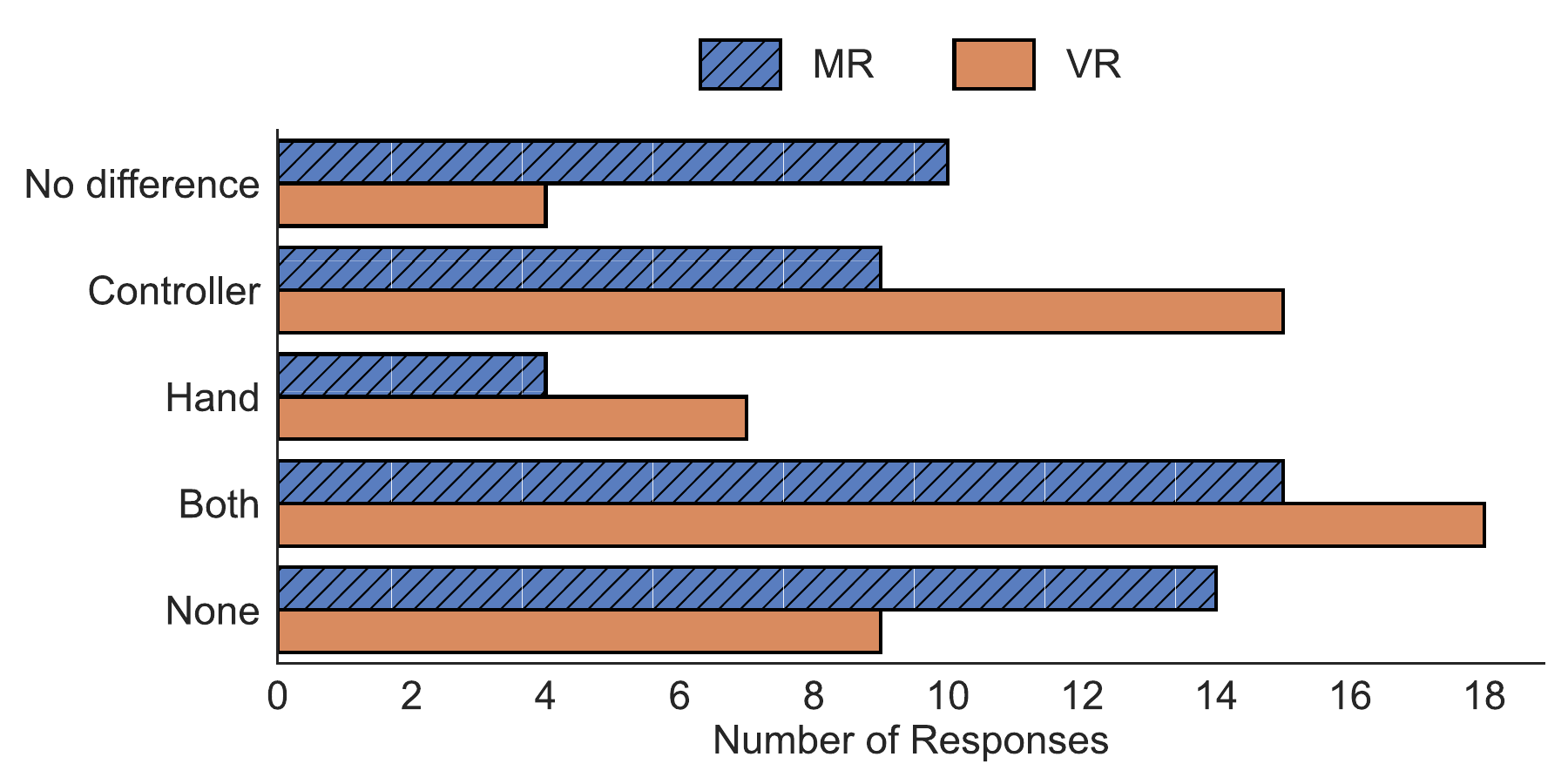}
    \caption{Number of participants who preferred each controller mode per XR mode (MR or VR).}
    \label{fig:controller-preference}
\end{figure}

Considering controller mode preference for VR, 18 participants preferred both hand and controller, and 15 preferred only the controller (Figure \ref{fig:controller-preference}). In the case of both hand and controller, participants indicated that it gave them superior information to assess the position and depth of the poke interactor and also allowed them to judge the angle of rotation of their hand easily. P4 stated, "Having both the hand and the controller as representation made it very easy to keep track of where the controller was at all times." These participants reported that the hand alone seemed unnatural and that it was difficult to judge the position of the poke interactor sphere alone. Those who preferred no representation felt the lack of visual information helped them concentrate on the position of the poke interactor more and felt more accurate (Figure~\ref{fig:controller-preference}). As P17 said, "The hands were distracting since they were a bit weird looking, being not my hand, and when it was only the dot, I had only the dot to see as exact reference instead of a controller as well." Participants who preferred the hand alone cited that its transparency and ability to mimic their hand position made it more realistic.

When considering controller representation preference for MR, participants reported an almost equal preference for both hand and controller representation, with 15 selecting it as a preference, and for no controller representation, which received 14 selections. Participants reported that they found the overlapped representation of either the controller or the hand was off-putting, as there was a mismatch with the real-world visual of their hand on the controller. P12 noted, "(The) virtual hand overlayed on top of the real hand felt weird because it was duplicated, (I) liked nothing (with just the dot) because you can see your hand and controller." Some participants also noted visible lag in the system, as mentioned earlier, for both conditions and felt that it impacted their performance (Figure~\ref{fig:controller-preference}). Notably, participants who preferred the combination of controller and hand, or "none" for the representation, did not report the lag. Participants who preferred no representation claimed they liked the reduced impact of the lagging visual, which showed just the tracking dot. 
\section{Discussion}

\subsection{Mixed vs. Virtual Reality} \label{mr_vs_vr}

Our first research question asked whether XR mode (VR vs. MR) influences selection task performance. Perhaps the most surprising result was that all measured dependent variables —selection time, effective throughput, depth deviation, and error rate —showed no significant differences between VR and MR modes. While a lack of statistical significance does not indicate equivalence, the means and standard deviations across metrics were also notably similar between MR and VR. This trend is further supported by our Fitts’ law analysis (see Figure~\ref{fig:fitts-all}). The linear regression results across VR and MR for all controller modes were very closely aligned, with the largest deviations observed when the virtual controller was present (intercept deviating by only 0.016 and slope by 0.009), which are still considered low. All models showed strong fits to the data ($R^2 > 0.791$). In addition, the visual plots in Figure~\ref{fig:fitts-vr} and Figure~\ref{fig:fitts-mr} exhibit very similar behaviour, with no controller representation (i.e., the None condition) resulting in a noticeably steeper slope. These results suggest that despite concerns about perceptual disruptions in MR, such as latency, occlusion, and other mismatched visuals \cite{drascic1996perceptual, kruijff2010perceptual}, these issues did not appear to affect selection performance in our study significantly. 

Despite these findings, our subjective results indicated that the change to the XR mode may impact a user's \emph{perceived} performance. Participants found changes to the controller representations less noticeable in the MR mode (Figure \ref{fig:xr-notice}) and reported visible effects of camera or tracking latency in the system. Although neither of these factors appeared to impact performance, they are worth noting, as user testing may reveal performance issues with the system that users imagine, despite evidence that those performance issues do not exist. 

\subsection{Controller Visualizations}

Our second research question asked whether the controller representation had any impact on selection performance. As expected, in comparing no controller representation (the None condition) to all other controller representations, selection time was higher (Figure \ref{fig:mt}) and effective throughput (Figure \ref{fig:tpe}) was lower. Depth deviation (Figure \ref{fig:depth}) was also higher with the None condition than all others, suggesting participants had greater difficulty correctly positioning the cursor at the correct depth without the visual context provided by  the controller model and / or hand surrogate.

Our Fitts' law regression models (Figure~\ref{fig:fitts-all}) further support these results:  for both VR and MR, the None condition regression lines had a steeper slope than all other representations (both, controller, and hand), all of which were largely parallel. This suggests that for more difficult selection tasks (i.e., higher $ID$), selection time increases faster without a controller representation than with \textit{any} representation is present. These findings are consistent with prior work, where representations like avatars have facilitated performance gains in selection tasks \cite{pan2019avatar, steed2016impact}. 

Our subjective results revealed some nuances when considering controller visualizations. While participants preferred the combined hand and controller representation in both VR and MR, a large portion of participants also preferred no controller representation in the MR mode (Figure \ref{fig:controller-preference}) despite the performance degradation exhibited in our results. Many of the subjective comments reported in Section~\ref{sec:preference} indicated that participants felt the controller and hand-only modes were unnatural. At the same time, the "none" condition followed the movements of their real hand grasping the controller. We would have expected to see a performance impact in MR specifically for these reasons based on previous research \cite{pan2019avatar, ponton_stretch_2024}. However, the results were similar to VR, suggesting that users may have different criteria for assessing personal performance in MR compared to VR.

\subsection{Depth Deviation}

While noted earlier in Section \ref{mr_vs_vr}, depth deviation did not significantly differ due to XR mode, the differences between MR and VR in the Hand and None controller conditions are notable and warrant further investigation into the components of depth. The lack of visual cues assisting participants in determining depth did have some impact, though not as intense as similar results from previous studies  \cite{jones2011-peripheral, swan2007egocentric}. Improvements in technology, particularly the visual fidelity of the pass-through camera feeds in recent hardware, may be contributing. It is also worth noting that the mean depth deviation in all conditions is high, with a mean of 1.27 cm. With an average target distance of 50 cm, this represents an error size of 0.03. While an approximately 1.25 cm depth deviation may seem minimal, it may still significantly impact real-world applications designed for free movement in a real environment. This is especially important today as MR modes become prevalent on VR devices, considering the relative simplicity of MR application development, particularly on modern Quest platforms. When designing for MR, developers must consider how this depth deviation may impact the safety of end-users and plan accordingly to provide feedback (visual, auditory, or haptic) to mitigate the effects of depth deviation in the real world. 
 
\subsection{Limitations and Future Work}

To understand the study's implications, we must acknowledge the limitations encountered. Based on the number of participants who reported seeing no changes to controller representation (see Figure \ref{fig:xr-notice}), the changes may have been too subtle between the conditions besides the no-controller condition. The changes in the XR environment were also not overly distinctive, with an almost exact recreation of the lab where the experiment took place (Figure \ref{fig:room-images}). Comparing more extreme changes in both representations may provide different results. Also, using standard embodiment measurements \cite{gonzalez-franco_avatar_2018} would allow for clearer assessments of the impressions participants may have on the controller representation in both XR modes.

Secondly, many users indicated experiencing the effects of latency in the MR conditions, where the virtual controller presentation lagged behind the video pass-through image of their real-life hand and controller movement. The specific impact of this latency was not considered. Although it did not appear to have a direct effect on any of the examined performance metrics, its repeated mention by several participants in the subjective feedback suggests that a thorough exploration of the overlapping latencies (tracking and video) may be insightful.

In the interest of building a framework for HCI in MR systems, this research can be extended in several directions. First, future work should measure the impact of conflicting latencies in pass-through-based MR devices and explore methods for compensating for them. Another direction is to investigate variations in controller representation, including differences in model fidelity, size, and realism, to better understand their influence on user perception and performance. The selection task itself can also be expanded upon by incorporating selection confirmation, changing the visual properties of targets, or exploring entirely different tasks altogether.
\section{Conclusion}

Our results suggest that selection performance in VR and MR may be fundamentally comparable. Converting from a VR environment to an MR environment does not appear to affect a user's ability to complete basic selection tasks. Similarly, the effects of controller representation on selection performance were consistent across XR modes; this is consistent with prior research and suggests that representations more closely resembling reality are associated with improved selection performance.

However, our subjective results indicate that user \emph{perception} of their selection performance differed between MR and VR, despite the lack of quantifiable differences in performance metrics. This suggests that selection interfaces for MR systems may require distinct design considerations from VR, particularly in terms of spatial awareness. Given the ease with which modern head-mounted displays and SDKs support the porting of VR applications to MR, this is a particularly timely area for future work.

\bibliographystyle{ACM-Reference-Format}
\bibliography{references}

@article{kovacs_perceptual_2008,
	title = {Perceptual influences on {Fitts}’ law},
	volume = {190},
	issn = {1432-1106},
	url = {https://doi.org/10.1007/s00221-008-1497-3},
	doi = {10.1007/s00221-008-1497-3},
	language = {en},
	number = {1},
	urldate = {2025-04-06},
	journal = {Experimental Brain Research},
	author = {Kovacs, A. J. and Buchanan, J. J. and Shea, C. H.},
	month = sep,
	year = {2008},
	pages = {99--103},
}

@article{kopper_human_2010,
	title = {A human motor behavior model for distal pointing tasks},
	volume = {68},
	issn = {1071-5819},
	url = {https://www.sciencedirect.com/science/article/pii/S1071581910000637},
	doi = {10.1016/j.ijhcs.2010.05.001},
	number = {10},
	urldate = {2025-03-25},
	journal = {International Journal of Human-Computer Studies},
	author = {Kopper, Regis and Bowman, Doug A. and Silva, Mara G. and McMahan, Ryan P.},
	month = oct,
	year = {2010},
	pages = {603--615},
}

@inproceedings{kim_effect_2025,
	address = {New York, NY, USA},
	series = {{CHI} {EA} '25},
	title = {The {Effect} of {Target} {Depth} on {Performance} of {Multi}-directional {Tapping} {Task} in {Virtual} {Reality}},
	isbn = {979-8-4007-1395-8},
	url = {https://dl.acm.org/doi/10.1145/3706599.3719893},
	doi = {10.1145/3706599.3719893},
	urldate = {2025-07-01},
	booktitle = {Proceedings of the {Extended} {Abstracts} of the {CHI} {Conference} on {Human} {Factors} in {Computing} {Systems}},
	publisher = {Association for Computing Machinery},
	author = {Kim, Haejun and Hong, Yuhwa and Yu, Jihae and Xiong, Shuping and Kim, Woojoo},
	year = {2025},
	pages = {1--8},
}

@inproceedings{hourcade_how_2012,
	address = {New York, NY, USA},
	series = {{CHI} '12},
	title = {How small can you go? analyzing the effect of visual angle in pointing tasks},
	isbn = {978-1-4503-1015-4},
	shorttitle = {How small can you go?},
	url = {https://dl.acm.org/doi/10.1145/2207676.2207706},
	doi = {10.1145/2207676.2207706},
	urldate = {2025-04-06},
	booktitle = {Proceedings of the {SIGCHI} {Conference} on {Human} {Factors} in {Computing} {Systems}},
	publisher = {Association for Computing Machinery},
	author = {Hourcade, Juan Pablo and Bullock-Rest, Natasha},
	month = may,
	year = {2012},
	pages = {213--216},
}

@article{fernandes_looking_2025,
	title = {Looking in {Depth}: {Targeting} by {Eye} and {Controller} {Input} for {Multi}-{Depth} {Target} {Placement}},
	volume = {41},
	issn = {1044-7318},
	shorttitle = {Looking in {Depth}},
	url = {https://doi.org/10.1080/10447318.2024.2401657},
	doi = {10.1080/10447318.2024.2401657},
	number = {13},
	urldate = {2025-06-27},
	journal = {International Journal of Human–Computer Interaction},
	author = {Fernandes, Ajoy S. and , T. Scott, Murdison and and Proulx, Michael J.},
	month = jul,
	year = {2025},
	note = {Publisher: Taylor \& Francis
\_eprint: https://doi.org/10.1080/10447318.2024.2401657},
	pages = {7952--7967},
}

@article{hoffman_vergenceaccommodation_2008,
	title = {Vergence–accommodation conflicts hinder visual performance and cause visual fatigue},
	volume = {8},
	issn = {1534-7362},
	url = {https://doi.org/10.1167/8.3.33},
	doi = {10.1167/8.3.33},
	number = {3},
	urldate = {2025-04-06},
	journal = {Journal of Vision},
	author = {Hoffman, David M. and Girshick, Ahna R. and Akeley, Kurt and Banks, Martin S.},
	month = mar,
	year = {2008},
	pages = {33},
}

@article{westermeierAssessingDepthPerception2024,
  title = {Assessing {{Depth Perception}} in {{VR}} and {{Video See-Through AR}}: {{A Comparison}} on {{Distance Judgment}}, {{Performance}}, and {{Preference}}},
  shorttitle = {Assessing {{Depth Perception}} in {{VR}} and {{Video See-Through AR}}},
  author = {Westermeier, Franziska and Br{\"u}bach, Larissa and Wienrich, Carolin and Latoschik, Marc Erich},
  year = {2024},
  month = may,
  journal = {IEEE Transactions on Visualization and Computer Graphics},
  volume = {30},
  number = {5},
  pages = {2140--2150},
  issn = {1941-0506},
  doi = {10.1109/TVCG.2024.3372061},
  urldate = {2025-07-09}
}

@inproceedings{liMixedRealityTunneling2022,
  title = {Mixed {{Reality Tunneling Effects}} for {{Stereoscopic Untethered Video-See-Through Head-Mounted Displays}}},
  booktitle = {2022 {{IEEE International Symposium}} on {{Mixed}} and {{Augmented Reality}} ({{ISMAR}})},
  author = {Li, Ke and Schmidt, Susanne and Bacher, Reinhard and Leemans, Wim and Steinicke, Frank},
  year = {2022},
  month = oct,
  pages = {44--53},
  issn = {1554-7868},
  doi = {10.1109/ISMAR55827.2022.00018},
  urldate = {2025-07-09}
}

@inproceedings{pointeckerBridgingGapRealities2022,
  title = {Bridging the {{Gap Across Realities}}: {{Visual Transitions Between Virtual}} and {{Augmented Reality}}},
  shorttitle = {Bridging the {{Gap Across Realities}}},
  booktitle = {2022 {{IEEE International Symposium}} on {{Mixed}} and {{Augmented Reality}} ({{ISMAR}})},
  author = {Pointecker, Fabian and Friedl, Judith and Schwajda, Daniel and Jetter, Hans-Christian and Anthes, Christoph},
  year = {2022},
  month = oct,
  pages = {827--836},
  issn = {1554-7868},
  doi = {10.1109/ISMAR55827.2022.00101},
  urldate = {2025-07-09}
}

@inproceedings{pointeckerRealVirtualExploring2024,
  title = {From {{Real}} to {{Virtual}}: {{Exploring Replica-Enhanced Environment Transitions}} along the {{Reality-Virtuality Continuum}}},
  shorttitle = {From {{Real}} to {{Virtual}}},
  booktitle = {Proceedings of the 2024 {{CHI Conference}} on {{Human Factors}} in {{Computing Systems}}},
  author = {Pointecker, Fabian and {Friedl-Knirsch}, Judith and Jetter, Hans-Christian and Anthes, Christoph},
  year = {2024},
  month = may,
  series = {{{CHI}} '24},
  pages = {1--13},
  publisher = {Association for Computing Machinery},
  address = {New York, NY, USA},
  doi = {10.1145/3613904.3642844},
  urldate = {2025-07-09},
  isbn = {979-8-4007-0330-0}
}

@inproceedings{otonoImTransformingEffects2023,
  title = {I'm {{Transforming}}! {{Effects}} of {{Visual Transitions}} to {{Change}} of {{Avatar}} on the {{Sense}} of {{Embodiment}} in {{AR}}},
  booktitle = {2023 {{IEEE Conference Virtual Reality}} and {{3D User Interfaces}} ({{VR}})},
  author = {Otono, Riku and Genay, Ad{\'e}la{\"i}de and {Perusqu{\'i}a-Hern{\'a}ndez}, Monica and Isoyama, Naoya and Uchiyama, Hideaki and Hachet, Martin and L{\'e}cuyer, Anatole and Kiyokawa, Kiyoshi},
  year = {2023},
  month = mar,
  pages = {83--93},
  issn = {2642-5254},
  doi = {10.1109/VR55154.2023.00024},
  urldate = {2025-07-09}
}

@inproceedings{feuchtnerExtendingBodyInteraction2017,
  title = {Extending the {{Body}} for {{Interaction}} with {{Reality}}},
  booktitle = {Proceedings of the 2017 {{CHI Conference}} on {{Human Factors}} in {{Computing Systems}}},
  author = {Feuchtner, Tiare and M{\"u}ller, J{\"o}rg},
  year = {2017},
  month = may,
  series = {{{CHI}} '17},
  pages = {5145--5157},
  publisher = {Association for Computing Machinery},
  address = {New York, NY, USA},
  doi = {10.1145/3025453.3025689},
  urldate = {2025-07-09},
  isbn = {978-1-4503-4655-9}
}

@article{genayBeingAvatarReal2022,
  title = {Being an {{Avatar}} ``for {{Real}}'': {{A Survey}} on {{Virtual Embodiment}} in {{Augmented Reality}}},
  shorttitle = {Being an {{Avatar}} ``for {{Real}}''},
  author = {Genay, Ad{\'e}la{\"i}de and L{\'e}cuyer, Anatole and Hachet, Martin},
  year = {2022},
  month = dec,
  journal = {IEEE Transactions on Visualization and Computer Graphics},
  volume = {28},
  number = {12},
  pages = {5071--5090},
  issn = {1941-0506},
  doi = {10.1109/TVCG.2021.3099290},
  urldate = {2025-07-09}
}

@inproceedings{otonoThirdPersonPerspectiveAvatar2022,
  title = {Third-{{Person Perspective Avatar Embodiment}} in {{Augmented Reality}}: {{Examining}} the {{Proteus Effect}} on {{Physical Performance}}},
  shorttitle = {Third-{{Person Perspective Avatar Embodiment}} in {{Augmented Reality}}},
  booktitle = {2022 {{IEEE Conference}} on {{Virtual Reality}} and {{3D User Interfaces Abstracts}} and {{Workshops}} ({{VRW}})},
  author = {Otono, Riku and Isoyama, Naoya and Uchiyama, Hideaki and Kiyokawa, Kiyoshi},
  year = {2022},
  month = mar,
  pages = {730--731},
  doi = {10.1109/VRW55335.2022.00216},
  urldate = {2025-07-09}
}

@inproceedings{ebrahimiInvestigatingEffectsAnthropomorphic2018,
  title = {Investigating the {{Effects}} of {{Anthropomorphic Fidelity}} of {{Self-Avatars}} on {{Near Field Depth Perception}} in {{Immersive Virtual Environments}}},
  booktitle = {2018 {{IEEE Conference}} on {{Virtual Reality}} and {{3D User Interfaces}} ({{VR}})},
  author = {Ebrahimi, Elham and Hartman, Leah S. and Robb, Andrew and Pagano, Christopher C. and Babu, Sabarish V.},
  year = {2018},
  month = mar,
  pages = {1--8},
  doi = {10.1109/VR.2018.8446539},
  urldate = {2025-07-09}
}

@inproceedings{lougiakisEffectsVirtualHand2020,
  title = {Effects of {{Virtual Hand Representation}} on {{Interaction}} and {{Embodiment}} in {{HMD-based Virtual Environments Using Controllers}}},
  booktitle = {2020 {{IEEE Conference}} on {{Virtual Reality}} and {{3D User Interfaces}} ({{VR}})},
  author = {Lougiakis, Christos and Katifori, Akrivi and Roussou, Maria and Ioannidis, Ioannis-Panagiotis},
  year = {2020},
  month = mar,
  pages = {510--518},
  issn = {2642-5254},
  doi = {10.1109/VR46266.2020.00072},
  urldate = {2025-07-09}
}

@inproceedings{mcmanusInfluenceAvatarSelf2011,
  title = {The Influence of Avatar (Self and Character) Animations on Distance Estimation, Object Interaction and Locomotion in Immersive Virtual Environments},
  booktitle = {Proceedings of the {{ACM SIGGRAPH Symposium}} on {{Applied Perception}} in {{Graphics}} and {{Visualization}}},
  author = {McManus, Erin A. and Bodenheimer, Bobby and Streuber, Stephan and {de la Rosa}, Stephan and B{\"u}lthoff, Heinrich H. and Mohler, Betty J.},
  year = {2011},
  month = aug,
  series = {{{APGV}} '11},
  pages = {37--44},
  publisher = {Association for Computing Machinery},
  address = {New York, NY, USA},
  doi = {10.1145/2077451.2077458},
  urldate = {2025-07-09},
  isbn = {978-1-4503-0889-2}
}

@article{seinfeldUserRepresentationsHumanComputer2021,
  title = {User {{Representations}} in {{Human-Computer Interaction}}},
  author = {Seinfeld, Sofia and Feuchtner, Tiare and Maselli, Antonella and M{\"u}ller, J{\"o}rg},
  year = {2021},
  month = oct,
  journal = {Human--Computer Interaction},
  volume = {36},
  number = {5-6},
  pages = {400--438},
  publisher = {Taylor \& Francis},
  issn = {0737-0024},
  doi = {10.1080/07370024.2020.1724790},
  urldate = {2025-07-09}
}

@inproceedings{johnson_exploring_2023,
	address = {Cham},
	title = {Exploring {Hand} {Tracking} and {Controller}-{Based} {Interactions} in a {VR} {Object} {Manipulation} {Task}},
	isbn = {978-3-031-48050-8},
	doi = {10.1007/978-3-031-48050-8\_5},
	language = {en},
	booktitle = {{HCI} {International} 2023 – {Late} {Breaking} {Papers}},
	publisher = {Springer Nature Switzerland},
	author = {Johnson, Cheryl I. and Fraulini, Nicholas W. and Peterson, Eric K. and Entinger, Jacob and Whitmer, Daphne E.},
	editor = {Chen, Jessie Y. C. and Fragomeni, Gino and Fang, Xiaowen},
	year = {2023},
	pages = {64--81},
}

@article{venkatakrishnan_how_2023,
	title = {How {Virtual} {Hand} {Representations} {Affect} the {Perceptions} of {Dynamic} {Affordances} in {Virtual} {Reality}},
	volume = {29},
	issn = {1941-0506},
	url = {https://ieeexplore.ieee.org/document/10049656},
	doi = {10.1109/TVCG.2023.3247041},
	number = {5},
	urldate = {2025-03-24},
	journal = {IEEE Transactions on Visualization and Computer Graphics},
	author = {Venkatakrishnan, Roshan and Venkatakrishnan, Rohith and Raveendranath, Balagopal and Pagano, Christopher C. and Robb, Andrew C. and Lin, Wen-Chieh and Babu, Sabarish V.},
	month = may,
	year = {2023},
	note = {Conference Name: IEEE Transactions on Visualization and Computer Graphics},
	pages = {2258--2268},
}

@inproceedings{hanashima_how_2022,
	address = {Berlin, Heidelberg},
	title = {How to {Elicit} {Ownership} and {Agency} for an {Avatar} {Presented} in the {Third}-{Person} {Perspective}: {The} {Effect} of {Visuo}-{Motor} and {Tactile} {Feedback}},
	isbn = {978-3-031-06508-8},
	shorttitle = {How to {Elicit} {Ownership} and {Agency} for an {Avatar} {Presented} in the {Third}-{Person} {Perspective}},
	url = {https://doi.org/10.1007/978-3-031-06509-5\_9},
	doi = {10.1007/978-3-031-06509-5\_9},
	urldate = {2025-02-13},
	booktitle = {Human {Interface} and the {Management} of {Information}: {Applications} in {Complex} {Technological} {Environments}: {Thematic} {Area}, {HIMI} 2022, {Held} as {Part} of the 24th {HCI} {International} {Conference}, {HCII} 2022, {Virtual} {Event}, {June} 26 – {July} 1, 2022, {Proceedings}, {Part} {II}},
	publisher = {Springer-Verlag},
	author = {Hanashima, Ryo and Ohyama, Junji},
	month = jun,
	year = {2022},
	pages = {111--130},
}

@article{hibbs_impact_2024,
	title = {Impact of virtual reality immersion on exercise performance and perceptions in young, middle-aged and older adults},
	volume = {19},
	issn = {1932-6203},
	url = {https://www.ncbi.nlm.nih.gov/pmc/articles/PMC11524493/},
	doi = {10.1371/journal.pone.0307683},
	number = {10},
	urldate = {2025-02-19},
	journal = {PLOS ONE},
	author = {Hibbs, Angela and Tempest, Gavin and Hettinga, Florentina and Barry, Gillian},
	month = oct,
	year = {2024},
	pmid = {39475865},
	pmcid = {PMC11524493},
	pages = {e0307683},
}

@inproceedings{lin_need_2016,
	address = {New York, NY, USA},
	series = {{SAP} '16},
	title = {Need a hand? how appearance affects the virtual hand illusion},
	isbn = {978-1-4503-4383-1},
	shorttitle = {Need a hand?},
	url = {https://dl.acm.org/doi/10.1145/2931002.2931006},
	doi = {10.1145/2931002.2931006},
	urldate = {2025-02-18},
	booktitle = {Proceedings of the {ACM} {Symposium} on {Applied} {Perception}},
	publisher = {Association for Computing Machinery},
	author = {Lin, Lorraine and Jörg, Sophie},
	month = jul,
	year = {2016},
	pages = {69--76},
}

@inproceedings{ponton_stretch_2024,
	address = {New York, NY, USA},
	series = {{CHI} '24},
	title = {Stretch your reach: {Studying} {Self}-{Avatar} and {Controller} {Misalignment} in {Virtual} {Reality} {Interaction}},
	isbn = {979-8-4007-0330-0},
	shorttitle = {Stretch your reach},
	url = {https://dl.acm.org/doi/10.1145/3613904.3642268},
	doi = {10.1145/3613904.3642268},
	urldate = {2025-02-18},
	booktitle = {Proceedings of the 2024 {CHI} {Conference} on {Human} {Factors} in {Computing} {Systems}},
	publisher = {Association for Computing Machinery},
	author = {Ponton, Jose Luis and Keshavarz, Reza and Beacco, Alejandro and Pelechano, Nuria},
	month = may,
	year = {2024},
	pages = {1--15},
}

@inproceedings{myersInteractingDistanceMeasuring2002,
  title = {Interacting at a Distance: Measuring the Performance of Laser Pointers and Other Devices},
  shorttitle = {Interacting at a Distance},
  booktitle = {Proceedings of the {{SIGCHI Conference}} on {{Human Factors}} in {{Computing Systems}}},
  author = {Myers, Brad A. and Bhatnagar, Rishi and Nichols, Jeffrey and Peck, Choon Hong and Kong, Dave and Miller, Robert and Long, A. Chris},
  year = {2002},
  month = apr,
  series = {{{CHI}} '02},
  pages = {33--40},
  publisher = {Association for Computing Machinery},
  address = {New York, NY, USA},
  doi = {10.1145/503376.503383},
  urldate = {2025-03-25},
  isbn = {978-1-58113-453-7}
}

@article{venkatakrishnanGiveMeHand2023,
  title = {Give {{Me}} a {{Hand}}: {{Improving}} the {{Effectiveness}} of {{Near-field Augmented Reality Interactions By Avatarizing Users}}' {{End Effectors}}},
  shorttitle = {Give {{Me}} a {{Hand}}},
  author = {Venkatakrishnan, Roshan and Venkatakrishnan, Rohith and Raveendranath, Balagopal and Pagano, Christopher C. and Robb, Andrew C. and Lin, Wen-Chieh and Babu, Sabarish V.},
  year = {2023},
  month = may,
  journal = {IEEE Transactions on Visualization and Computer Graphics},
  volume = {29},
  number = {5},
  pages = {2412--2422},
  issn = {1941-0506},
  doi = {10.1109/TVCG.2023.3247105},
  urldate = {2025-07-04}
}

@misc{metaMetaQuestMR2025,
  title = {Meta {{Quest MR}}, {{VR Headsets}} \& {{Accessories}}},
  author = {Meta},
  year = {2025},
  urldate = {2025-03-07},
  howpublished = {https://www.meta.com/ca/quest},
  langid = {english}
}

@book{castronova2003theory,
  title={Theory of the Avatar},
  author={Castronova, E.},
  series={CESifo working papers},
  url={https://books.google.ca/books?id=nHAeNAEACAAJ},
  year={2003},
  publisher={CES}
}

@article{fitts1954information,
  title={The information capacity of the human motor system in controlling the amplitude of movement.},
  author={Fitts, Paul M},
  journal={Journal of experimental psychology},
  volume={47},
  number={6},
  pages={381},
  year={1954},
  publisher={American Psychological Association}
}

@INPROCEEDINGS{heidicker2017influence,
  author={Heidicker, Paul and Langbehn, Eike and Steinicke, Frank},
  booktitle={2017 IEEE Symposium on 3D User Interfaces (3DUI)}, 
  title={Influence of avatar appearance on presence in social VR}, 
  year={2017},
  volume={},
  number={},
  pages={233-234},
  doi={10.1109/3DUI.2017.7893357}}

@article{hillis2004slant,
    author = {Hillis, James M. and Watt, Simon J. and Landy, Michael S. and Banks, Martin S.},
    title = {Slant from texture and disparity cues: Optimal cue combination},
    journal = {Journal of Vision},
    volume = {4},
    number = {12},
    pages = {1-1},
    year = {2004},
    month = {12},
    issn = {1534-7362},
    doi = {10.1167/4.12.1},
    url = {https://doi.org/10.1167/4.12.1},
    eprint = {https://arvojournals.org/arvo/content\_public/journal/jov/933504/jov-4-12-1.pdf},
}

@inproceedings{knierim2018physical,
author = {Knierim, Pascal and Schwind, Valentin and Feit, Anna Maria and Nieuwenhuizen, Florian and Henze, Niels},
title = {Physical Keyboards in Virtual Reality: Analysis of Typing Performance and Effects of Avatar Hands},
year = {2018},
isbn = {9781450356206},
publisher = {Association for Computing Machinery},
address = {New York, NY, USA},
url = {https://doi.org/10.1145/3173574.3173919},
doi = {10.1145/3173574.3173919},
booktitle = {Proceedings of the 2018 CHI Conference on Human Factors in Computing Systems},
pages = {1–9},
numpages = {9},
location = {Montreal QC, Canada},
series = {CHI '18}
}

@ARTICLE{kourtesis2022action,
  author={Kourtesis, Panagiotis and Vizcay, Sebastian and Marchal, Maud and Pacchierotti, Claudio and Argelaguet, Ferran},
  journal={IEEE Transactions on Visualization and Computer Graphics}, 
  title={Action-Specific Perception \& Performance on a Fitts's Law Task in Virtual Reality: The Role of Haptic Feedback}, 
  year={2022},
  volume={28},
  number={11},
  pages={3715-3726},
  doi={10.1109/TVCG.2022.3203003}}

@inproceedings{kress2017optical,
author = {Bernard C. Kress and William J. Cummings},
title = {{Optical architecture of HoloLens mixed reality headset}},
volume = {10335},
booktitle = {Digital Optical Technologies 2017},
editor = {Bernard C. Kress and Peter Schelkens},
organization = {International Society for Optics and Photonics},
publisher = {SPIE},
pages = {103350K},
year = {2017},
doi = {10.1117/12.2270017},
URL = {https://doi.org/10.1117/12.2270017}
}

@INPROCEEDINGS{kruijff2010perceptual,
  author={Kruijff, Ernst and Swan, J. Edward and Feiner, Steven},
  booktitle={2010 IEEE International Symposium on Mixed and Augmented Reality}, 
  title={Perceptual issues in augmented reality revisited}, 
  year={2010},
  volume={},
  number={},
  pages={3-12},
  doi={10.1109/ISMAR.2010.5643530}}

@article{macKenzie1992fitts,
author = {I. Scott MacKenzie},
title = {Fitts' Law as a Research and Design Tool in Human-Computer Interaction},
journal = {Human–Computer Interaction},
volume = {7},
number = {1},
pages = {91--139},
year = {1992},
publisher = {Taylor \& Francis},
doi = {10.1207/s15327051hci0701\_3},
URL = {https://doi.org/10.1207/s15327051hci0701\_3},
eprint = {https://doi.org/10.1207/s15327051hci0701\_3}
}

@inproceedings{mackenzie1992extending,
author = {MacKenzie, I. Scott and Buxton, William},
title = {Extending Fitts' law to two-dimensional tasks},
year = {1992},
isbn = {0897915135},
publisher = {Association for Computing Machinery},
address = {New York, NY, USA},
url = {https://doi.org/10.1145/142750.142794},
doi = {10.1145/142750.142794},
booktitle = {Proceedings of the SIGCHI Conference on Human Factors in Computing Systems},
pages = {219–226},
numpages = {8},
location = {Monterey, California, USA},
series = {CHI '92}
}

@inproceedings{mackenzie2008fitts,
author = {MacKenzie, I. Scott and Isokoski, Poika},
title = {Fitts' throughput and the speed-accuracy tradeoff},
year = {2008},
isbn = {9781605580111},
publisher = {Association for Computing Machinery},
address = {New York, NY, USA},
url = {https://doi.org/10.1145/1357054.1357308},
doi = {10.1145/1357054.1357308},
booktitle = {Proceedings of the SIGCHI Conference on Human Factors in Computing Systems},
pages = {1633–1636},
numpages = {4},
location = {Florence, Italy},
series = {CHI '08}
}

@inproceedings{mason2001reaching,
author = {Mason, Andrea H. and Walji, Masuma A. and Lee, Elaine J. and MacKenzie, Christine L.},
title = {Reaching movements to augmented and graphic objects in virtual environments},
year = {2001},
isbn = {1581133278},
publisher = {Association for Computing Machinery},
address = {New York, NY, USA},
url = {https://doi.org/10.1145/365024.365308},
doi = {10.1145/365024.365308},
booktitle = {Proceedings of the SIGCHI Conference on Human Factors in Computing Systems},
pages = {426–433},
numpages = {8},
location = {Seattle, Washington, USA},
series = {CHI '01}
}

@article{mcgee1997fitts,
author = {Mike McGee and Brian Amento and Patrick Brooks and Hope Harley},
title ={Fitts and VR: Evaluating Display and Input Devices with Fitts' Law},

journal = {Proceedings of the Human Factors and Ergonomics Society Annual Meeting},
volume = {41},
number = {2},
pages = {1259-1262},
year = {1997},
doi = {10.1177/1071181397041002119},
URL = {
        https://doi.org/10.1177/1071181397041002119
},
eprint = { 
    
        https://doi.org/10.1177/1071181397041002119
}
}

@article{milgram1994taxonomy,
  title={A taxonomy of mixed reality visual displays},
  author={Milgram, Paul and Kishino, Fumio},
  journal={IEICE TRANSACTIONS on Information and Systems},
  volume={77},
  number={12},
  pages={1321--1329},
  year={1994},
  publisher={The Institute of Electronics, Information and Communication Engineers}
}

@inproceedings{pan2019avatar,
author = {Pan, Ye and Steed, Anthony},
title = {Avatar Type Affects Performance of Cognitive Tasks in Virtual Reality},
year = {2019},
isbn = {9781450370011},
publisher = {Association for Computing Machinery},
address = {New York, NY, USA},
url = {https://doi.org/10.1145/3359996.3364270},
doi = {10.1145/3359996.3364270},
booktitle = {Proceedings of the 25th ACM Symposium on Virtual Reality Software and Technology},
articleno = {6},
numpages = {4},
location = {Parramatta, NSW, Australia},
series = {VRST '19}
}

@ARTICLE{skarbez2021revisiting,

AUTHOR={Skarbez, Richard  and Smith, Missie  and Whitton, Mary C. },

TITLE={Revisiting Milgram and Kishino's Reality-Virtuality Continuum},

JOURNAL={Frontiers in Virtual Reality},

VOLUME={2},

YEAR={2021},

URL={https://www.frontiersin.org/journals/virtual-reality/articles/10.3389/frvir.2021.647997},

DOI={10.3389/frvir.2021.647997},

ISSN={2673-4192}}

@inproceedings{steed2016impact,
  title={The impact of a self-avatar on cognitive load in immersive virtual reality},
  author={Steed, Anthony and Pan, Ye and Zisch, Fiona and Steptoe, William},
  booktitle={2016 IEEE virtual reality (VR)},
  pages={67--76},
  year={2016},
  organization={IEEE}
}

@inproceedings{ware1995dynamic,
  title={Dynamic stereo displays},
  author={Ware, Colin},
  booktitle={Proceedings of the SIGCHI conference on human factors in computing systems},
  pages={310--316},
  year={1995}
}

@article{weidner2023systematic,
  title={A systematic review on the visualization of avatars and agents in ar \& vr displayed using head-mounted displays},
  author={Weidner, Florian and Boettcher, Gerd and Arboleda, Stephanie Arevalo and Diao, Chenyao and Sinani, Luljeta and Kunert, Christian and Gerhardt, Christoph and Broll, Wolfgang and Raake, Alexander},
  journal={IEEE Transactions on Visualization and Computer Graphics},
  volume={29},
  number={5},
  pages={2596--2606},
  year={2023},
  publisher={IEEE}
}

@book{ware2000information,
  title={Information visualization: perception for design},
  author={Ware, Colin},
  year={2000},
  publisher={Morgan Kaufmann}
}

@article{iso2012human, title={Ergonomics of human-system interaction — Part 411: Evaluation methods for the design of physical input devices (ISO/TS 9241-411:2012(en))}, author={International Organization for Standardization}, year={2012}}

@inproceedings{art2011,
author = {Wobbrock, Jacob O. and Findlater, Leah and Gergle, Darren and Higgins, James J.},
title = {The aligned rank transform for nonparametric factorial analyses using only anova procedures},
year = {2011},
isbn = {9781450302289},
publisher = {Association for Computing Machinery},
address = {New York, NY, USA},
url = {https://doi.org/10.1145/1978942.1978963},
doi = {10.1145/1978942.1978963},
booktitle = {Proceedings of the SIGCHI Conference on Human Factors in Computing Systems},
pages = {143–146},
numpages = {4},
location = {Vancouver, BC, Canada},
series = {CHI '11}
}

@inproceedings{elkin2021-artc,
author = {Elkin, Lisa A. and Kay, Matthew and Higgins, James J. and Wobbrock, Jacob O.},
title = {An Aligned Rank Transform Procedure for Multifactor Contrast Tests},
year = {2021},
isbn = {9781450386357},
publisher = {Association for Computing Machinery},
address = {New York, NY, USA},
url = {https://doi.org/10.1145/3472749.3474784},
doi = {10.1145/3472749.3474784},
booktitle = {The 34th Annual ACM Symposium on User Interface Software and Technology},
pages = {754–768},
numpages = {15},
location = {Virtual Event, USA},
series = {UIST '21}
}

@article{swan2007egocentric,
  title={Egocentric depth judgments in optical, see-through augmented reality},
  author={Swan, J Edward and Jones, Adam and Kolstad, Eric and Livingston, Mark A and Smallman, Harvey S},
  journal={IEEE transactions on visualization and computer graphics},
  volume={13},
  number={3},
  pages={429--442},
  year={2007},
  publisher={IEEE}
}

@article{thompson2004-quality,
author = {Thompson, William B. and Willemsen, Peter and Gooch, Amy A. and Creem-Regehr, Sarah H. and Loomis, Jack M. and Beall, Andrew C.},
title = {Does the quality of the computer graphics matter when judging distances in visually immersive environments},
year = {2004},
issue_date = {October 2004},
publisher = {MIT Press},
address = {Cambridge, MA, USA},
volume = {13},
number = {5},
issn = {1054-7460},
url = {https://doi.org/10.1162/1054746042545292},
doi = {10.1162/1054746042545292},
journal = {Presence: Teleoper. Virtual Environ.},
month = oct,
pages = {560–571},
numpages = {12}
}

@inproceedings{drascic1996perceptual,
  title={Perceptual issues in augmented reality},
  author={Drascic, David and Milgram, Paul},
  booktitle={Stereoscopic displays and virtual reality systems III},
  volume={2653},
  pages={123--134},
  year={1996},
  organization={Spie}
}

@inproceedings{jones2011-peripheral,
author = {Jones, J. Adam and Swan, J. Edward and Singh, Gurjot and Ellis, Stephen R.},
title = {Peripheral visual information and its effect on distance judgments in virtual and augmented environments},
year = {2011},
isbn = {9781450308892},
publisher = {Association for Computing Machinery},
address = {New York, NY, USA},
url = {https://doi.org/10.1145/2077451.2077457},
doi = {10.1145/2077451.2077457},
booktitle = {Proceedings of the ACM SIGGRAPH Symposium on Applied Perception in Graphics and Visualization},
pages = {29–36},
numpages = {8},
location = {Toulouse, France},
series = {APGV '11}
}

@INPROCEEDINGS{jones2008-effects,
  author={Jones, Adam and Swan, J. Edward and Singh, Gurjot and Kolstad, Eric},
  booktitle={2008 IEEE Virtual Reality Conference}, 
  title={The Effects of Virtual Reality, Augmented Reality, and Motion Parallax on Egocentric Depth Perception}, 
  year={2008},
  volume={},
  number={},
  pages={267-268},
  doi={10.1109/VR.2008.4480794}}

@article{buck2018-comparison,
author = {Buck, Lauren E. and Young, Mary K. and Bodenheimer, Bobby},
title = {A Comparison of Distance Estimation in HMD-Based Virtual Environments with Different HMD-Based Conditions},
year = {2018},
issue_date = {July 2018},
publisher = {Association for Computing Machinery},
address = {New York, NY, USA},
volume = {15},
number = {3},
issn = {1544-3558},
url = {https://doi.org/10.1145/3196885},
doi = {10.1145/3196885},
journal = {ACM Trans. Appl. Percept.},
month = jul,
articleno = {21},
numpages = {15}
}

@article{witmer_measuring_1998,
	title = {Measuring {Presence} in {Virtual} {Environments}: {A} {Presence} {Questionnaire}},
	volume = {7},
	shorttitle = {Measuring {Presence} in {Virtual} {Environments}},
	url = {https://doi.org/10.1162/105474698565686},
	doi = {10.1162/105474698565686},
	number = {3},
	urldate = {2023-02-27},
	journal = {Presence: Teleoperators and Virtual Environments},
	author = {Witmer, Bob G. and Singer, Michael J.},
	month = jun,
	year = {1998},
	pages = {225--240},
}

@article{kilteni_sense_2012,
	title = {The {Sense} of {Embodiment} in {Virtual} {Reality}},
	volume = {21},
	url = {https://doi.org/10.1162/PRES\_a\_00124},
	doi = {10.1162/PRES\_a\_00124},
	number = {4},
	urldate = {2025-04-11},
	journal = {Presence: Teleoperators and Virtual Environments},
	author = {Kilteni, Konstantina and Groten, Raphaela and Slater, Mel},
	month = nov,
	year = {2012},
	pages = {373--387},
}

@article{slater_body_1994,
	title = {Body centred interaction in immersive virtual environments},
	volume = {1},
	number = {1994},
	journal = {Artificial life and virtual reality},
	author = {Slater, Mel and Usoh, Martin},
	year = {1994},
	pages = {125--148},
}

@inproceedings{kabash_prince_1995,
    author = {Kabbash, Paul and Buxton, William A. S.},
    title = {The “prince” technique: Fitts' law and selection using area cursors},
    year = {1995},
    isbn = {0201847051},
    publisher = {ACM Press/Addison-Wesley Publishing Co.},
    address = {USA},
    url = {https://doi.org/10.1145/223904.223939},
    doi = {10.1145/223904.223939},
    booktitle = {Proceedings of the SIGCHI Conference on Human Factors in Computing Systems},
    pages = {273–279},
    numpages = {7},
    location = {Denver, Colorado, USA},
    series = {CHI '95}
}

@inproceedings{demarbre_investigating_2024,
	location = {New York, {NY}, {USA}},
	title = {Investigating Presence Across Rendering Style and Ratio of Virtual to Real Content in Mixed Reality},
	isbn = {979-8-4007-1088-9},
	url = {https://doi.org/10.1145/3677386.3682098},
	doi = {10.1145/3677386.3682098},
	series = {{SUI} '24},
	pages = {1--8},
	booktitle = {Proceedings of the 2024 {ACM} Symposium on Spatial User Interaction},
	publisher = {Association for Computing Machinery},
	author = {{DeMarbre}, Eric and Henderson, Jay and Teather, Robert J},
	date = {2024-10-07},
}

@inproceedings{amini_systematic_2025,
	location = {New York, {NY}, {USA}},
	title = {A Systematic Review of Fitts' Law in 3D Extended Reality},
	isbn = {979-8-4007-1394-1},
	url = {https://dl.acm.org/doi/10.1145/3706598.3713623},
	doi = {10.1145/3706598.3713623},
	series = {{CHI} '25},
	pages = {1--25},
	booktitle = {Proceedings of the 2025 {CHI} Conference on Human Factors in Computing Systems},
	publisher = {Association for Computing Machinery},
	author = {Amini, Mohammadreza and Stuerzlinger, Wolfgang and Teather, Robert J and Batmaz, Anil Ufuk},
	urldate = {2025-06-22},
	date = {2025-04-25},
}

@article{gonzalez-franco_avatar_2018,
	title = {Avatar Embodiment. Towards a Standardized Questionnaire},
	volume = {5},
	issn = {2296-9144},
	url = {https://www.frontiersin.org/journals/robotics-and-ai/articles/10.3389/frobt.2018.00074/full},
	doi = {10.3389/frobt.2018.00074},
	journaltitle = {Frontiers in Robotics and {AI}},
	shortjournal = {Front. Robot. {AI}},
	author = {Gonzalez-Franco, Mar and Peck, Tabitha C.},
	urldate = {2025-09-09},
	date = {2018-06-22},
	note = {Publisher: Frontiers},
}
\end{document}